\documentclass[12pt]{article}
\usepackage{epsfig}
\usepackage{amsfonts}
\usepackage[tbtags]{amsmath} 
\usepackage{amsthm} 
\usepackage{times}
\usepackage{amsmath}%
\usepackage{amssymb}%
\usepackage{graphicx}
\usepackage{epstopdf}
\usepackage{rotating}
\usepackage{booktabs}
\usepackage{caption}
\usepackage[mathscr]{euscript}
\usepackage{lscape}
\usepackage{subcaption}
\usepackage{mathrsfs}
\usepackage{subcaption}
\usepackage{times}
\usepackage{rotating}
\usepackage{float}
\usepackage{afterpage}
\usepackage{url}
\usepackage{xcolor,colortbl}
\usepackage{geometry}
\usepackage{multirow}
\usepackage{graphics}
\usepackage{dsfont}
\usepackage{bm}
\usepackage{enumitem}
\usepackage{bibentry}

\usepackage{float}
\usepackage{adjustbox}

\definecolor{Gray}{gray}{0.85}
\definecolor{LightCyan}{rgb}{0.88,1,1}
\usepackage{natbib}
\usepackage{setspace,xr}
\usepackage[implicit=false]{hyperref}

\newtheorem{remark}{REMARK}[section]

\newtheorem{guess}{ASSUMPTION}[section]

\def\captionof#1#2{{\def\@captype{#1}#2}}
\makeatletter
\renewcommand\paragraph{\@startsection{paragraph}{4}{\z@}%
	{-3.25ex\@plus -1ex \@minus -.2ex}%
	{1.5ex \@plus .2ex}%
	{\normalfont\normalsize\bfseries}}
\makeatother

\usepackage{tikz}
\usepackage{quantikz}
\usetikzlibrary{trees}
\usetikzlibrary{positioning}
\usetikzlibrary{arrows.meta}
\usetikzlibrary{positioning}
\usetikzlibrary{decorations.markings}
\usetikzlibrary{decorations.text}
\usetikzlibrary{matrix}
\usepackage{float}
\usepackage{footmisc}

\usepackage{adjustbox}

\usepackage{float}

\usepackage{fancyhdr}
\usepackage{xr}
\begin{document}
	
\begin{titlepage}
	\centering
	{\Large
		\renewcommand{\thefootnote}{\fnsymbol{footnote}}
		On Quantum Ambiguity and Potential Exponential Computational Speed-Ups to Solving Dynamic Asset Pricing Models\footnote{Manuscript Received November 2024, Revised June 2025}}
	
	\vspace{1cm}
	
	\renewcommand{\thefootnote}{\arabic{footnote}} \setcounter{footnote}{0}
	Eric Ghysels\footnote{We thank the referees for their invaluable comments which helped us substantially improve our paper and to Jes\'us Fern\'andez-Villaverde for insightful suggestions to expand the analysis of our paper. The first author acknowledges the financial support from an IBM Global University Program Academic Award and an IonQ QLAB Global Users Program grant. While conducting this research we benefited from conversations with Andrii Babii, Lars Hansen, Isaiah Hull, Hamed Mohammadbagherpoor, and George Tauchen. We also thank Jieyao Wang for research assistance. Conference participants of the 2024 SoFiE Annual meetings in Rio de Janeiro, Brazil and seminar participants at the Kenan-Flagler Business School, the Research Triangle Quantum Webinar series and the University of Chicago provided helpful feedback.} \\
\medskip
	{\normalsize
		University of North Carolina Chapel Hill, U.S.A, Rethinc.Labs Kenan Institute,\\
		CEPR and Kenan-Flagler Business School} \\
	\medskip
	Jack Morgan \\
	\medskip
	{\normalsize
		University of North Carolina Chapel Hill, U.S.A, Rethinc.Labs Kenan Institute}
	\bigskip
	\begin{abstract}
		\renewcommand{\thefootnote}{\arabic{footnote}}
		\noindent We formulate quantum computing solutions to a large class of dynamic nonlinear asset pricing models using algorithms, in theory exponentially more efficient than classical ones, which leverage the quantum properties of superposition, entanglement and interference.  The equilibrium asset pricing solution is a quantum state. We use quantum decision-theoretic foundations of ambiguity and model/parameter uncertainty to deal with model selection.
	\end{abstract} 
	\vfill
\end{titlepage}

	\renewcommand{\thefootnote}{\arabic{footnote}}
	\setcounter{footnote}{1}

\thispagestyle{empty}
\setcounter{page}{0}	
\newpage

\newpage
\section{Introduction}

Quantum computing (QC) is a paradigm shift in computer science that aims to use the principles of quantum mechanics to solve computational problems that are  beyond the capabilities of classical computers.
A quantum computer is a device designed to execute quantum algorithms using a physical quantum system as a qubit. The term ``quantum computer" encompasses both quantum annealers designed for optimization problems, and universal or ``gate based'' quantum computers. The algorithms that we will discuss in this work are designed for universal quantum computers. When executing an algorithm on a quantum processor, the resulting state is not an exact recreation of the state promised by the algorithm due to hardware errors where the physical system undergoes an uncontrolled and unintentional change. This hardware noise grows with the number of steps and qubits required by the algorithm.  Quantum computing is poised to reshape the landscape of modern finance.  At its core, quantum computing is a new type of computing based on the rules of quantum physics. Traditional computers use {\it bits} — units of information that are either 0 or 1 — to perform calculations. Quantum computers, however, use {\it quantum bits}, or {\it qubits}, which can exist as 0, 1, or any combination of both at once, thanks to a phenomenon called {\it superposition}. This allows quantum computers to process a vast number of possibilities at the same time, making them powerful tools for solving problems that are too complex for today’s most advanced supercomputers. Another key concept is {\it entanglement}, which describes a unique relationship between qubits. When qubits are entangled, the state of one qubit is directly related to the state of another, even if they’re far apart. This allows quantum systems to perform coordinated computations that can capture complex relationships and patterns more efficiently than classical systems. The current state of quantum computing is promising but still in a nascent stage. In the foreseeable future, quantum hardware will become mainstream, in particular in a hybrid combination with classical computing. \cite{adedoyin2018quantum} provide a great introduction and survey of quantum computing for readers without a physics background. We aim to review the core essentials required to see the advantages of QC for the class of models in this paper. 

\medskip

An arbitrary quantum state $\Psi$ can be represented by the superposition of two basis vectors e.g. $\Psi_0 = a_0 \ket{0} + b_0 \ket{1}$ where $|a_0|^2 + |b _0|^2$ =1, where $a_0$ and $b_0$ are complex numbers.\footnote{We use here and throughout the paper, the so-called Dirac notation. The Dirac bra-ket notation uses $\ket{u_i}$ (called a ket) for the complex column vector $u_i$ and $\bra{u_i}$ (a bra) is written for its adjoint, the row vector containing the complex conjugates of its elements, with the complex conjugate written as $\overline{u}_i$. The quadratic form $\overline{u}_i A u_i$ is then written as $\bra{u_i} A \ket{u_i}$. The notation allows one to distinguish numbers from matrices, as in $\braket{u_i}{u_i}$ versus $\ket{u_i} \bra{u_i}$, and to specify vectors through labels or descriptions as in $\ket{\text{Model}\, 1}$.  See \cite{nielsen_chuang_2010} for a standard textbook reference or the Appendix to \cite{morrell2021step} which provides a quick summary focused on the concepts used in our paper.} When a measurement is made in the computational basis, $\Psi_0$ will collapse to either $\ket{0}$ with probability $|a|^2$ or $\ket{1}$ with probability $|b_0|^2$. A single qubit can also be in a classical mixture of pure states like $\Psi_3 = a_3 \Psi_1 + b_3 \Psi_2$ where $\Psi_1$ and $\Psi_2$ are their own superposed states. In certain situations the state of one qubit is dependent on the state of a separate qubit. These qubits are ``entangled'' and their states can only be represented as a single, four dimensional vector. Any operation on one qubit will instantaneously affect its entangled partner. By adding one qubit, we double the size of the vector space a quantum computer can use to store information. Quantum algorithms leverage quantum entanglement to solve problems in an entirely different manner than their classical counterparts, sometimes with (exponentially) fewer resources.

\medskip

Some promising applications in finance include: (a) quantum Monte Carlo simulation where a potential of quadratic speed-up has been pursued, see e.g.\ \cite{woerner2019quantum} (and see \cite{skavysh2023quantum} for a general discussion), (b) QC also produces potential quadratic speed-ups in computing quantiles and are therefore applied to credit risk models, see \cite{egger2020credit} and \cite{ghysels2023quantum}, (c) option pricing, see e.g.\  \cite{stamatopoulos2020option} for Black-Scholes type models with exotic option pricing. See \cite{kaneko2022quantum}, \cite{vazquez2021efficient}, \cite{ghysels2023quantum}, for option pricing models with stochastic volatility, among others, (d) portfolio optimization, see e.g.\ \cite{rebentrost2024quantum}. There are also a number of recent literature surveys, including \cite{orus2019quantum} and \cite{herman2022survey}, in addition to the textbook by \cite{hull2020quantum}.\footnote{\label{footnoteblog} There are also several industry blogs, including: IBM Quantum, which includes open source implementations such as the {\it Qiskit Finance Module}, see \cite{javadi2024quantum}, and D-Wave, the latter explaining how quantum annealers (as opposed to gate-based quantum computers) can be used for financial optimization problems, with real examples and code structure. Also useful is the Quantum Computing Zoo website, see \cite{QuantumAlgorithmZoo}.}

 We propose a new class of numerical solutions to dynamic asset pricing models using QC algorithms. As the equilibrium solution comes in the form of a quantum state, we use a theoretical and practical framework for econometricians who cope with model selection in the face of misspecification and ambiguity,  assembling elements from (a) quantum computing and measurement, (b) statistical decision theory and (c) quantum models of decision-making.

\medskip

We are interested in the potential opportunities quantum computing can bring to advance fundamental research in asset pricing. More specifically, we  are interested in solving an equation of interest in dynamic asset pricing models:
\begin{equation}
\nu(x_t)  =  \int \psi(x_{t+1},x_t) \nu(x_{t+1}) f(x_{t+1}|x_t) dx_{t+1} + g(x_t). \label{eq:inteq2}
\end{equation}
The equation,
also known as Fredholm equations of the second type, involves functions $\psi(\cdot,\cdot)$ and $g(\cdot)$ which depend on economic determinants, including the stochastic discount factor (SDF). We are interested in characterizing $\nu(\cdot)$, the solution to a dynamic asset pricing problem, such as the equilibribum price-dividend ratio.\footnote{We deliberately start from a discrete time setting while keeping in mind that continuous time asset pricing models either have discrete time analogues to equation 
(\ref{eq:inteq2}) or some discretization provides is a workable approximation.}

\medskip

We can rewrite equation (\ref{eq:inteq2}) as follows $\nu$ = $\mathcal{T} \left[ \nu \right]$ + $g$,
where $\mathcal{T} \left[ \nu \right]$ is the operator defined by the integral term, which under suitable regularity conditions is invertible, i.e.\ the operator $\left[ I - \mathcal{T} \right]^{-1}$ exists, where $I$ denotes the identity operator.\footnote{We defer to Section  \ref{sec:AP} the formal discussion regarding the conditions for a proper integral and associated operator  $\nu$ = $\mathcal{T} \left[ \nu \right]$ + $g$ in equation (\ref{eq:inteq2}).} This yields the  solution $\nu$ = $\left[ I - \mathcal{T} \right]^{-1}g$. The quadrature approach of \cite{tauchen1991quadrature} amounts to proposing an approximate solution $\nu_N$ = $\left[ I - \mathcal{T}_N \right]^{-1}b_N$ using $\mathcal{T}_N$ instead of $\mathcal{T}$ with the former a quadrature-based discrete version close to the latter. The approximation error diminishes as $N$ $\rightarrow$ $\infty$. 

\medskip



Practical implementations often require inverting a high-dimensional matrix.
In this paper we propose to use QC algorithms to handle these high-dimensional computations. With classical computers the time complexity grows at best linearly in $N$. More precisely, a classical computer using the conjugate gradient method requires $O(N s \kappa \log(1/\varepsilon))$ running time, where $s$ is a measure of sparsity, $\kappa$ a conditioning number and $\varepsilon$ the accuracy of the approximation.\footnote{The condition number of the matrix is defined as the ratio of the smallest and the largest eigenvalue and sparsity $s$ measures the max non-zero entries per row or column. Time complexity $O(\cdot)$ denotes an upper bound on the worst case of a problem. Inverting an $N$-dimensional square symmetric matrix via elementary methods (such as Gaussian elimination) theoretically takes $O(N^3)$ amount of time, i.e.\ has a polynomial rate of the third order. More sophisticated algorithms, using the conjugate gradient method achieve linear rates.} The QC algorithm has a running time complexity of $O(\log_2(N) s^2 \kappa^2/\varepsilon)$,  an exponential speed-up in the size of the system (i.e.\ QC algorithms
have a worse runtime than conjugate gradient descent in terms of the error and condition number, but are
exponentially better in terms of the dimension of the system).
One crucial caveat to keep in mind is that QC hardware is not as reliable as classical computing, but the fidelity of QC is steadily improving as innovations in hardware are implemented.  

\medskip

Solutions to asset pricing models, using the methods proposed in this paper, are represented by quantum states. This comes with its own challenges but also offers new exciting opportunities. In particular, we need to draw attention to the fact that measuring output generated by QC algorithms is not as straightforward as it is with classical computers. When a measurement is performed on a quantum system, it ``collapses'' to a specific state with certain probabilities.  
In physics, measurement refers to physical instruments used in lab experiments to measure quantum states. Moreover, in physics almost all data is experimental and the famous von Neumann measurement model assumes that - appart from measurement noise - the data is aligned with quantum states being generated. We use a novel quantum decision-theoretic approach to ambiguity and model/parameter uncertainty, building on \cite{ghysels2024measure}, and apply it to  asset pricing model selection. 

\medskip

Our paper takes an approach that is different from the recent work by \cite{fernandez2022dynamic}. They  use a specific type of hardware, commercialized by D-Wave, known as a quantum annealer, specialized to performs combinatorial optimization. The important innovation in the paper by \cite{fernandez2022dynamic} is to formulate a dynamic programming problem as a combinatorial optimization problem and show how their approach is able to recover value and policy functions. Moreover, their algorithm is 
not limited by fundamental scaling bottlenecks, and they showcase its implementation using a small (but non-trivial) problem of solving a real business cycle model on existing quantum hardware. We take a different approach. First, we focus exclusively on dynamic asset pricing models which require the solution to integral equations of the type appearing in equation (\ref{eq:inteq2}). Second, we explore solution methods that can be implemented on any gate-based quantum hardware. This offers a wider range of potential implementations on current and future quantum computing hardware potentially achieving the theoretical exponential speedups. It is important to emphasize, however, that implementing the methods proposed in our paper are challenging with the currently available QC hardware. Hence, our paper is not (yet) about showing QC speed-ups currently achievable. Instead, we expect the steady developments in hardware to make the research methods we propose feasible in the near future. We therefore use classical methods to mimic QC hardware as a proof of concept instead.



\medskip

The paper is structured as follows. In Section \ref{sec:AP} we start with describing the class of asset pricing models and summarize the quadrature approach of \cite{tauchen1991quadrature} to solving such models. 
In Section \ref{sec:QC} we present the QC algorithms. Section  \ref{sec:implementation} reports on an empirical asset pricing implementation. Conclusions appear in Section \ref{sec:concl}.

\medskip

\noindent {\bf Notation:} The Hadamard product between two same-sized matrices $A = \left[a_{ij}\right]_{i,j} \in \mathbb{R}^{N_1\times N_2}$ and $B \left[b_{ij}\right]_{i,j} \in \mathbb{R}^{N_1\times N_2}$ is denoted by $A\circ B\in \mathbb{R}^{N_1\times N_2}$ corresponds to the element-wise matrix product, i.e.\ $A\circ B$ := $\left[a_{ij}b_{ij}\right]_{i,j}$. 
The computational part of the paper adopts the  \cite{dirac1939new} bra-ket notation used in quantum mechanics, as well as the quantum computation and information literature (see \cite{nielsen_chuang_2010} for a standard textbook reference or the Appendix to \cite{morrell2021step} which provides a quick summary focused on the concepts used in our paper).\footnote{The Dirac bra-ket notation uses $\ket{u_i}$ (called a ket) for the complex column vector $u_i$ and $\bra{u_i}$ (a bra) is written for its adjoint, the row vector containing the complex conjugates of its elements, with the complex conjugate written as $\overline{u}_i$. The quadratic form $\overline{u}_i A u_i$ is then written as $\bra{u_i} A \ket{u_i}$. The notation allows
one to distinguish numbers from matrices, as in $\braket{u_i}{u_i}$ versus $\ket{u_i} \bra{u_i}$, and to specify vectors through labels or descriptions as in $\ket{\text{Model}\, 1}$.} The basic building block is a Hilbert space $\mathbb{H}$, which is a complete normed vector space over $\mathbb{C}$ with inner product denoted
$\braket \cdot \cdot$ : $\mathbb{H}$ $\times$ $\mathbb{H} \rightarrow \mathbb{C}$.  Of particular interest to us is the tensor product space $\bigotimes_{i=1}^n \mathbb{H}$ and orthonormal basis which will be denoted $\ket{u_i}$ for $i$ = 0, $\ldots$, $n-1$.

\setcounter{equation}{0}
\section{Dynamic Asset Pricing Models \label{sec:AP}}

We start with a probability space, namely, a triple ($\Omega, \mathcal{F}, \mathcal{P}$) where $\Omega$ is a collection of discrete time ($t$ $\in$ $\mathbb{N}$) infinite sequences in $\mathbb{R}^{n_y}$, $\mathcal{F}$ is the smallest sigma algebra of events in $\Omega$ and $\mathcal{P}$
assigns probabilities to
events. The state of the economy is described by the $n_y$-dimensional stationary stochastic process $\{y_t \in \mathbb{R}^{n_y} : t = 1, \ldots, \infty\}$. Also of interest is the sigma filtration $\mathcal{F}_t$ and associated $\mathcal{P}_t$ pertaining to events up to time $t$.
We focus on a single asset with ex-dividend price $p_{t}$ at time time $t$  and future dividend stream $\{d_{t+k}, k = 1, \ldots, \infty\}$. Denote the price-dividend ratio by $\nu_{t}$ = $p_{t}/d_{t}$. Given a representative agent's time $t$ stochastic discount factor (SDF) $m(y_{t+1}, y_t)$, we have:
\begin{equation}
	\label{eq:mrs}
	\nu_{t} = \mathbb{E}_t \left[ (1 + \nu_{t+1}) h_{t+1} m( y_{t+1}, y_t)\right]
\end{equation}
where $h_{t+1}$ = $d_{t+1}/d_{t}$ = $h(y_{t+1})$ is the dividend growth and $\mathbb{E}_t[\cdot]$ the conditional expectation given the $\mathcal{F}_t$ filtration. We restrict our attention to stationary Markov processes with conditional distribution of $y_t$ given its  past $x_{t-1}$ = $\{y_{t-1}\}$  given by $f(y_t|x_{t-1})$. We can write equation (\ref{eq:mrs}) in integral form as follows:
\begin{eqnarray}
	\label{appeq:mrsint}
	\nu(x) & = & \int [1 + \nu(y)] \psi(y,x) f(y|x) dy \notag \\
	\underset{\nu}{\underbrace{\nu(x)}}  & = & \underset{\mathcal{T}[\nu]}{\underbrace{\int \nu(y) \psi(y,x) \frac{f(y|x)}{\omega(y)} \omega(y) dy}} + \underset{g}{\underbrace{\int  \psi(y,x) \frac{f(y|x)}{\omega(y)} \omega(y) 	 dy }} \notag \\
		\nu & = & \mathcal{T} \left[ \nu \right] + g,
\end{eqnarray}
where $\nu(x):$ $\mathbb{R}^{n_y}$ $\rightarrow$ $\mathbb{R}$ is the price-dividend ratio as a function of the current state, while $\psi(y,x)$ = $h(y)$ $\times$ $m(y,x)$ with $h(y)$ the dividend growth as function of the future state and $m(y,x)$  the SDF. 
In Online Appendix Section \ref{appsubsec:quad} we review the quadrature approximation approach to solving dynamic asset pricing models introduced by \cite{tauchen1991quadrature}. 
A quadrature rule can be viewed as a discrete probability model that approximates the density $\omega$, and in the case of Gauss quadrature rules these approximations are close to minimum norm rules,  yielding for $\psi_{jk}$ = $\psi(\bar{y}_k,\bar{y}_j):$
\begin{equation}
\label{eq:linquadeq}
\bar{\nu}_{Nj}  =  \sum_{k=1}^N \left[1 + \bar{\nu}_{Nk} \right] \psi_{jk} \pi^N_{jk}  =  \sum_{k=1}^N  \left[\bar{\nu}_{Nk}  \right] \psi_{jk} \pi^N_{jk} + \sum_{k=1}^N \left[1 \right] \psi_{jk} \pi^N_{jk} \quad j = 1, \ldots, N
\end{equation}
where $\pi^N_{jk}$ = $\pi_k^N(\bar{y}_j)$ for $\pi_k^N(x)$ = $[f(\bar{y}_k|x)/(s(x) \omega(\bar{y}_k))] w_k$ and $s(x)$ = $\sum_{j=1}^{N} [f(\bar{y}_j|x)/\omega(\bar{y}_j)] w_i$. Therefore:
\begin{equation}
\label{eq:linquadfinal}
\underset{\bar{\nu}_{N}}{\underbrace{
		\left[
		\begin{array}{c}
		\bar{\nu}_{N1} \\
		\vdots \\
		\bar{\nu}_{Nj} \\
		\vdots \\
		\bar{\nu}_{NN}
		\end{array}
		\right]}}	 =  
\underset{\mathcal{T}_N \bar{\nu}_{N}}{\underbrace{\left[
		\begin{array}{cccc}
		\psi_{11} \pi^N_{11} & \ldots & &  \psi_{1N} \pi^N_{1N}	 \\
		& & & \\
		\vdots &  \ddots & & \vdots \\
		& &   & \\
		\psi_{N1} \pi^N_{N1} & \ldots & & \psi_{NN} \pi^N_{NN}
		\end{array}
		\right]
		\left[
		\begin{array}{c}
		\bar{\nu}_{N1} \\
		\vdots \\
		\bar{\nu}_{Nj} \\
		\vdots \\
		\bar{\nu}_{NN}
		\end{array}
		\right]}}
+ \underset{b_N}{\underbrace{
		\left[
		\begin{array}{c}
		\sum_{k=1}^N \psi_{1k} \pi^N_{1k} \\
		\vdots \\
		\sum_{k=1}^N	 \psi_{jk} \pi^N_{jk} \\
		\vdots \\
		\sum_{k=1}^N	 \psi_{Nk} \pi^N_{Nk}
		\end{array}
		\right]}}.
\end{equation}
To proceed, recall that  $\psi(y,x)$ = $h(y)$ $\times$ $m(y,x)$, which motivates defining the following matrices $\Psi_N$ := $\left[\psi_{ij}\right]_{i,j =1, \ldots, N}$, $\mathcal{M}_N$ := $\left[m_{ij}\right]_{i,j =1, \ldots, N}$, $\Pi_N$ := $\left[\pi_{ij}\right]_{i,j =1, \ldots, N}$, $\mathcal{H}_N$ = $\left[[h_i]_{i= 1, \ldots, N} \times \mathbf{1}_N^\prime\right]$ where $\mathbf{1}_N$ is a $N$ $\times$ 1 vector of ones, and finally $\Psi_N$ = $\mathcal{H}_N \circ \mathcal{M}_N$.  Under Assumption \ref{appassum:inverse} the following is an approximate solution to the fundamental pricing equation (\ref{eq:inteq2}):
\begin{equation}
	\label{eq:discreteinversion1}
	\bar{\nu}_{N} =  [I_N - \Psi_N \circ \Pi_N]^{-1} b_N \quad \text{ or equivalently } \quad  \bar{\nu}_{N} =  [I_N - \mathcal{H}_N \circ \mathcal{M}_N \circ \Pi_N]^{-1} b_N
\end{equation}
which can be viewed as the solution to the asset pricing equations where the law of motion of the state vector is a discrete Markov chain with the $N$ quadrature abscissa $\bar{y}_j$ as states and transition probabilities $\pi^N_{jk}$ = Pr($y_t$ =  $\bar{y}_k$ $\vert$ $y_{t-1}$ = $\bar{y}_j$). 
\medskip

Finally, we introduce some generic notation for equation (\ref{eq:discreteinversion1}) and a variation of it more suitable for quantum computing. Starting with the former, we will use the following generic notation:
\begin{equation}
	\label{eq:finalquadeq}
	\mathcal{A}_N	\bar{\nu}_{N} =   b_N \quad \text{ with } \mathcal{A}_N = [I_N - \mathcal{T}_N] = [I_N -  \mathcal{H}_N \circ \mathcal{M}_N \circ \Pi_N]\quad \text{ and therefore } \bar{\nu}_{N} = \mathcal{A}_N^{-1}b_N.
\end{equation}
Quantum mechanics involve unitary operator applications to unit-norm vectors in Hilbert spaces. What that in mind, we will rewrite $b_N$ as:
$b_N$  $\equiv$ diag$(\sqrt{N}\sum_{k=1}^N \psi_{1k} \pi^N_{1k}, \ldots, 
		\sqrt{N}\sum_{k=1}^N	 \psi_{Nk} \pi^N_{Nk}) \times \iota_N$ 
	 $\equiv$  $\mathcal{B}_N \iota_N,$
with $\mathcal{B}_N$ = diag($\sqrt{N}\sum_{k=1}^N \psi_{1k} \pi^N_{1k}, \ldots, 
\sqrt{N}\sum_{k=1}^N	 \psi_{Nk} \pi^N_{Nk}$) and $\iota_N$ is a unit-norm vector in a $N$-dimensional space. We can rewrite equation (\ref{eq:finalquadeq}) as:
\begin{equation}
	\label{eq:finalquadequnitnorm}
	(\mathcal{B}^{-1}_N\mathcal{A}_N)	\bar{\nu}_{N} =   \iota_N \quad \text{ and therefore } \bar{\nu}_{N} = \mathcal{C}_N^{-1}\iota_N \quad \text{ with } \mathcal{C}_N = (\mathcal{B}^{-1}_N\mathcal{A}_N).
\end{equation}
Note that $\mathcal{C}_N$ = $\mathcal{B}^{-1}_N\mathcal{A}_N$ encodes all the information about the model and we can therefore focus on $\mathcal{C}_N$ when we study multiple models.

\setcounter{equation}{0}
\setcounter{table}{0}
\section{Quantum Computational Solutions for Asset Pricing Models \label{sec:QC}}

Quantum algorithms for solving linear systems of equations are named HHL, after the three authors Aram Harrow, Avinatan Hassidim and Seth Lloyd, who introduced them (see \cite{harrow2009quantum}). The HHL algorithm leverages the quantum properties of superposition and entanglement to solve certain types of linear systems more efficiently than classical algorithms, in theory with an exponential speedup. The HHL algorithm focuses on solving systems of linear equations of the form $A \ket x$ = $\ket b$, where $A$ is an $n$ $\times$ $n$ matrix and using the Dirac notation for $\ket x$ and $\ket b$, which means that they will refer to quantum states, as further explained later.

\medskip

The HHL algorithm involves five main components, namely
(1) state preparation - which prepares a quantum state that encodes the input vector $b$, (2) quantum phase estimation (QPE) to extract the eigenvalues of the matrix $A$, (3) 
ancilla bit rotation, (4) inverse quantum phase estimation
(IQPE), and finally (5) measurement to extract (a function of) the solution vector $x$. 
More specifically, the first step of the algorithm  is to write $\ket b$ in terms of the eigenvectors $\ket{a_i}$ corresponding to the eigenvalues $\lambda_i$ of $A$, namely:
$\ket b$ = $\sum_{i=0}^{2^{n_b}-1} \beta_i \ket{a_i}$,
we then find the solution as: $\ket x$ = $\sum_{i=0}^{2^{n_b}-1} \lambda_i^{-1} \beta_i \ket{a_i}$.\footnote{A typical QC algorithm relies on superpositions, representing a combination of multiple states simultaneously. $\ket b$ = $\sum_{i=0}^{2^{n_b}-1} \beta_i \ket{a_i}$, is an example of superposition involving combinations of $\ket{a_i}$. In general the $\beta_i$ can be complex-valued, an issue further discussed in the next section.} So, the second step is finding the eigenvectors and eigenvalues of $A$. This is achieved by so-called Quantum Phase Estimation (QPE) which consists of controlled rotations applied to quibts and a Inverse Quantum Fourier Transform (IQFT). While this yields a solution, it is one with the eigenvectors of $A$ as the basis, while we want the solution to be in terms of a standard orthonormal basis, i.e.\ $\ket{u_i}$. Therefore we need to undo the IQFT, or put differently applying QFT, which is next step in the algorithm. The final step is measurement, i.e.\ reading the output of the solution $\ket x$ we obtained. Quantum measurement is a fundamental concept in quantum mechanics that refers to the process of obtaining information about the properties of a quantum system. When a measurement is made, the system collapses into one of its possible states, known as an eigenstate, with a certain probability. 

\medskip

First, the intuition regarding the exponential speed-up.
Classical computers use bits to represent information, quantum computers use qubits (quantum bits), which can exist in multiple states at once to perform calculations. Let's  look at approximate solution $\nu_N$ = $\left[ I - \mathcal{T}_N \right]^{-1}b_N$. It takes $n_b$ qubits to store the information in a vector such as $b$ of size $N$ = $2^{n_b}$ (assuming $n_b$ is integer). The vector $b$ is translated to a quantum state and linear algebraic operations, such as applying $\left[ I - \mathcal{T}_N \right]^{-1}$, amount to manipulation of the information in the $n_b$ qubits using quantum mechanics principles which are formally described as unitary operators defined on $n_b$-dimensional tensor products of Hilbert spaces. Hence, we stay within the $n_b$-dimensional setup going through motions which amount to matrix inversion. These motions have a time complexity proportional to $\log_2 N$. While this is certainly an oversimplification, it is intuitively why there is a potential exponential speed-up. 

\medskip

Second we discuss why it matters. Suppose we have a four dimensional state variable $y_t$. One can think of monthly data, say aggregate consumption growth, and we take current month and three lags which yields the four dimensional state process, i.e.\ $n_y$ = 4. In addition, suppose we take a five point quadrature rule. That means $N$ = $5^4$ = $625$ whereas $\log_2 N$ $<$ 10. When we add another series with four lags and move to a nine point quadrature rule we have $N$ = $9^8$ = $43,046,721$ whereas $\log_2 N$ $<$ 26. Even though this is only a small-scale example with two economic driving processes each having three lags, the numerical computations grow quickly beyond our reach or are too costly with classical computing.
When we add model uncertainty, the above arguments are even more of critical importance. For example, \cite{hansen2007beliefs} considers four univariate AR model 
parameter configurations or submodels, where a submodel is a collection of states
for which there is no chance of leaving that
collection. Each of the four models have a 100-point discretitzed Markov chain, adding up to a 400 state Markov chain
to approximate a model selection problem or
estimation problem for investors. While this is a simple example, it clearly illustrates how robustness concerns amplify the computational burden, in the case of classical computers it grows linear and multiplicative in the number of models, i.e.\ $N_M \times N$, where $N_M$ is the number of models. For QC algorithms the computational time complexity is additive on a $\text{log}_2$ scale in $N_M$ since $\log_2 (N_M \times N)$ = $\log_2 N_M$ + $\log_2 N$.\footnote{This argument is based on the fact that one can stack models, such as  $\{\bar{\nu}_{N_D}^m(\theta_m) :  m = 1, \ldots, N_M\}$, in a large dimensional system. See e.g.\ \cite{hansen2007beliefs} for an example.}

\medskip

Assuming that the state $\ket{b}$ is given as an input, the runtime of HHL scales with  $O(\log_2(N) s^2 \kappa^2/\varepsilon)$ where $s$ is a measure of sparsity, $\kappa$ a conditioning number and $\varepsilon$ the accuracy of the approximation. The exponential speedup HHL offers with respect to the system size is offset by the runtime dependence on sparsity and conditioning. Therefore, HHL is a promising approach only for applications when $A$ is sparse and well conditioned. We explore these characteristics for the proposed asset pricing models in Online Appendix Section \ref{appsec:hhl-feasibility}. Quantum alternatives to HHL with better dependence on $s$ and $\kappa$ are available for problems that do not fit this criteria. \cite{wossnig2018quantum}  introduced a method using quantum singular value estimation (QSVE) that is independent of $s$ for dense matrices. For poorly conditioned matrices, \cite{ambainis2010variable} leverage the novel variable time amplitude amplification subroutine to create a quantum linear system algorithm that scales almost linearly with $\kappa$. In addition, variations of HHL have been developed by \cite{yalovetzky2022nisq} and \cite{MORGAN2025130181} which are more amenable to noisy hardware, and offer potential advantages for poorly conditioned problems. \cite{dervovic2018quantumlinearsystemsalgorithms} provide and excellent review of the various options.\footnote{\cite{zheng2024earlyinvestigationhhlquantum} summarize the limitations of current hardware specifically pertaining to linear algebra problems, and conclude that the largest matrix accurately inverted on a real gate based processor is 4x4.}

\medskip

The runtimes of all candidate algorithms are based on the assumption that the vector $\ket{b}$ is provided as an input. The process of preparing the state $b$ is an open problem that affects many quantum applications. The naive approach scales with $\mathcal{O}(N)$ which would dominate the runtime of the full algorithm. Quantum random access memory (QRAM) could reduce this complexity to $O(\log_2(N))$ (see \cite{PhysRevLett.100.160501}). Machine learning algorithms like the one developed by \cite{han2025enqodefastamplitudeembedding} could reduce the state preparation time, thus securing the exponential speedup of the complete quantum linear algebra algorithm in practice. For a review of the many approches being pursued to tackle this problem, see \cite{herman2022survey}. 

\medskip

To solve the matrix inversion problem in equation (\ref{eq:finalquadeq}) with one of these quantum approaches, we first need to convert the problem into a quantum linear system problem of the form $A_N \ket x$ = $\ket b$, where $A$ is an $N$ $\times$ $N$ Hermitian matrix and $\ket{x}$ and $\ket{b}$ are $n$ dimensional quantum states. In equation (\ref{eq:finalquadeq}), the inverted matrix $B$ is not necessarily Hermitian, so we define:
\begin{equation}
	\label{eq:hermitianC}
	A_{N} = 
	\begin{pmatrix} 0 & \mathcal{A}_N \\ \overline{\mathcal{A}}_N & 0 \end{pmatrix}
\end{equation}

where $\overline{\mathcal{A}}_N$ is the convex conjugate of $\mathcal{A}_N$ and we can solve the equation $Ax$ = $b$ where 
\[
b = \begin{pmatrix} b_N \\ 0 \end{pmatrix} \qquad x = \begin{pmatrix}
	0 \\ \bar{\nu}_{N}  \end{pmatrix}
\]
where $\bar{\nu}_{N}$ is the solution to the asset pricing equations appearing in equations (\ref{eq:discreteinversion1}) and (\ref{eq:finalquadeq}) which is our main objective. Note also that equation (\ref{eq:hermitianC}) implies that $A$ is of dimension 2 $\times$ $N$, with $N$ the number of quadrature abscissa used in the discrete approximation formula. We suppress the dependence on $N$ for $A$, $b$ and $x$ in order to simplify notation, although we will revisit the impact of $N$ at a later stage. Moreover, in our applications the matrices $\mathcal{A}_N$ and vectors  $b_N$ have real-valued entries, such that we can replace Hermitian with symmetric and complex conjugate with transpose, but for the sake of generality we stick with the QC jargon and setup. That being said, when it comes to the actual implementation we will take advantage of the fact that there are no imaginary parts to the inputs. Moreover, we work with unit vectors, and therefore we scale the inputs accordingly.

\medskip

So far, we described $\ket x$ as the generic solution to a quantum linear algebra problem. At this point we need to remind ourselves that we are computing the solution to a dynamic asset pricing problem and that the inversion yields a quantum state best described as $\ket{\bar{\nu}_{N}(\theta)}$. To simplify notation from now on we will drop the $N$ subscript, and write $\ket{\bar{\nu}(\theta)}$ instead of $\ket{\bar{\nu}_{N}(\theta)}$. The model solution encodes the equilibrium price-dividends sorted from low to high. 

\setcounter{equation}{0}
\setcounter{table}{0}
\setcounter{figure}{0}
\section{Data, Models and Measurement Operators \label{sec:ambig}}

The measurement process is probabilistic, meaning that the outcome of the measurement is not deterministic but instead determined by properties of the quantum state and the measurement operator.  \cite{harrow2009quantum} suggest that when applying their algorithm, one should consider not looking at the actual solution $\ket x$,  but rather compute as output  $\bra x \mathbb{A} \ket x$, in the spirit of the von Neumann model of measurement in quantum mechanics, for some judiciously chosen operator $\mathbb{A}$.\footnote{As forcefully argued by \cite{aaronson2015read}, it would be, among other things, counterproductive to read out all the individual elements of $\ket x$.} The question is which $\mathbb{A}$ should one pick.

\subsection{Data and Pricing Errors and Benchmark Model \label{subsec:datapricingerrors}}

We first introduce a quantum state pertaining to the data, which we write as $\ket d$. Assuming finite support for price-dividend ratios (an assumption also made by \cite{tauchen1991quadrature}), we can divide up the interval between the empirical min and max of observed price-dividend ratios with the min as the first and the max as the $N^{th}$ point and thus construct an (equally-spaced) empirical counterpart for the purpose of model comparisons with the data.\footnote{Note that the equally spaced grid points may not coincide with the quadrature abscissa. Although we do not cover the case $N$ $\rightarrow$ $\infty$, it is reasonable to assume those differences will vanish with finer grid points.} From this $N$-dimensional vector we can create the $n_\nu$-dimensional quantum state $\ket d$. It is worth reminding ourselves at this point that all quantum states are unit-norm. This means that the construction of $\ket d$ also involves scaling to have a unit norm representation. Next, we impose the following technical condition:
\begin{guess}
	\label{assum:nonzerolength}
	For any model $\ket{\bar{\nu}(\theta)}$ the following holds: $\ket{\bar{\nu}(\theta)}$ $\neq$ $\ket d$.
\end{guess}
\noindent The above assumption implies that there is no perfect fit. At this stage we do not discuss the reasons. It may be model specification error, but it may also be the result of approximation error in  $\ket{\bar{\nu}(\theta)}$ and/or sampling error in $\ket d$. 

\medskip

In quantum computing one distinguishes pure state, i.e.\ in our case a single asset pricing model, from mixed states, a statistical ensemble of multiple models. We will first focus on pure states and thus a single model and in particular on the {\it pricing error} pure state $\vert d - \bar{\nu}(\theta) \rangle.$ 
For any measurement operator $\mathbb{A}$, applied to the pure state $\vert d - \bar{\nu}(\theta) \rangle$ we compute expectation values using a well-known formula in quantum mechanics for the expectation value of an observable.\footnote{\label{footnoteBorn} To be more precise, the Born rule for pure states $\ket \phi$ and a measurement operator $\mathbb{A}$ with eigenvectors $\ket{a_i}$ corresponding to eigenvalues $\lambda_{i}$ (assuming for simplicity that they are unique) states that the probability distribution for the measurement outcomes $\mathbb{P}(\lambda_{i})$ = $|\langle a_i | \phi \rangle|^2$ and the expectation value is therefore $\sum_i \lambda_{i} |\langle a_i | \phi \rangle|^2$ = $\bra{\phi} \mathbb{A} \ket{\phi}$, see e.g.\ \cite{scherer2019mathematics} p.\ 30. The Born rule was originally stated as a postulate and later proven for Hilbert spaces of dimension 3 or more as a theorem (known as the Gleason theorem). For a mixed state $\rho$  and measurement operator the expectation is tr($\mathbb{A} \rho$) and the Born rule states $\mathbb{P}(\lambda_{i})$ = tr($P_i \rho$). A special case of a mixed state is a pure one for which the expectation value is tr($\mathbb{A} P_\phi$). In equation (\ref{eq:utility}) we use interchangeably the formulas:
	$\langle d - \bar{\nu}(\theta) \vert \mathbb{A}\vert d - \bar{\nu}(\theta) \rangle$ and tr($\mathbb{A}P_{d - \bar{\nu}(\theta)}$) in the case of pure states.} Namely we have the expectation value:
\begin{equation}
	\label{eq:utility}
	\langle d - \bar{\nu}(\theta) \vert \mathbb{A}\vert d - \bar{\nu}(\theta) \rangle = \text{tr}(\mathbb{A}P_{d - \bar{\nu}(\theta)}) \quad \text{ with } P_{d - \bar{\nu}} = \vert d - \bar{\nu}(\theta) \rangle \langle d - \bar{\nu}(\theta) \vert,
\end{equation}
where $P_{d - \bar{\nu}}$ is the projection operator on the space spanned by $\ket{d - \bar{\nu}(\theta)}$. One choice for $\mathbb{A}$ is to look only at the data, namely $\mathbb{A}$ = $P_{d},$ i.e.\ projection on the data quantum state $\ket{d}$, yielding the following:
\begin{eqnarray}
	\label{eq:CvM}
	\langle d - \bar{\nu}(\theta) \vert  P_{d} \vert d - \bar{\nu}(\theta) \rangle  & = & 	\langle d - \bar{\nu}(\theta) \vert P_{d} P_{d}  \vert d - \bar{\nu}(\theta) \rangle \quad P_d \text{ idempotent} \\
	& = & 	 \langle e(\theta) \ket{e(\theta)} \qquad \text{ where } \ket{d - P_d \bar{\nu}(\theta)} = \ket{e(\theta) }\notag \\
	& = & \|\ket{e(\theta)} \|^2 \notag 
\end{eqnarray}
which is reminiscent of Cramer-von Mises statistics because $\ket d$ can be interpreted as a quantum state reflecting the empirical cumulative density and $\ket{\bar{\nu}(\theta)}$ the model counterpart.   

\medskip

To construct other measurement operators, it will be useful to express the data quantum state $\ket d$ in terms of its $N$-dimensional orthonormal basis  $\ket{u_i}$ for $i$ = 0, $\ldots$, $N-1,$ with $\ket d$ = $\sum_i \delta_i^d \ket{u_i}.$

\subsection{Benchmark and Tail Event Operators \label{subsec:benchmarks}}

In asset pricing, as in many other fields there is a benchmark model, represented by $\ket{\bar{\nu}_B},$ the quantum state representing the solution of the benchmark model. This yields the second measurement operator we will consider, which is a classical mixture of benchmark model and data:
\begin{eqnarray}
	\label{eq:secondAAcase}
	\mathbb{A}  =  (1 - p) P_d + p P_B & &   \\
	\langle d - \bar{\nu}(\theta) \vert \mathbb{A}\vert d - \bar{\nu}(\theta) \rangle & =	 &  (1 - p) \text{tr}(P_{d}P_{d - \bar{\nu}(\theta)}) + p \text{tr}(P_BP_{d - \bar{\nu}(\theta)}), \notag
\end{eqnarray}
where one might be ambivalent about the data and puts theory, via the benchmark model, ahead of measurement as p moves closer to one.\footnote{There are many other possibilities for measurement based on a benchmark model, such as phase operators such as: $\mathbb{A}$ = $\ket{\bar{\nu}_B} \bra{\bar{\nu}_B}$ - $\ket{\bar{\nu}_B^\perp} \bra{\bar{\nu}_B^\perp},$  as suggested by one of the referees. \label{footnote:phase} More generally, one can construct $\mathbb{A}$ as a unitary operator that applies a phase $e^{i\theta}$ to states in $P_B$ = $\ket{\bar{\nu}_B} \bra{\bar{\nu}_B},$ and does nothing to states in the orthogonal space $P_{B^\perp}$ = $\ket{\bar{\nu}_B^\perp} \bra{\bar{\nu}_B^\perp}:$	$\mathbb{A}$ = $e^{i\theta} P_B + P_{B^\perp}$ where $\mathbb{A}$ applies a phase shift $\theta$ to any component in the subspace projected by $P_B$.} 

\medskip

Next we consider measurement operators based on extreme tail events, as they may be useful in evaluation rare disaster arguments in asset pricing. For this we focus on two specific quantum states, namely $\ket d,$ i.e.\ the data quantum state and $\vert d - \bar{\nu}(\theta) \rangle,$ the pricing error associated with a particular model. Recall that  $\ket d$ = $\sum_i \delta_i^d \ket{u_i},$ and therefore we can isolate tail events by picking only basis states $\ket{u_i}$ from the set $B_d,$ standing for bad outcomes according to the data. We can then think of operators of the form $\mathbb{A}$ =
$\sum_{i \in B_d} \delta_i^d \ket{u_i} \bra{u_i}.$  They could apply to $\ket d,$  $\vert d - \bar{\nu}(\theta) \rangle,$ or the state associated with a particular model, i.e.\ the pure quantum state $\vert \bar{\nu}(\theta) \rangle.$ In the empirical implementation we will focus on the worst data outcome which we will associate with the basis quantum state $\ket{u_0}$ and consider the $\mathbb{A}$ = $\ket{u_0} \bra{u_0}$ = $P_0.$

\subsection{Measurement Operators and Quantum Ambiguity \label{subsec:measurementambiguity}}

Last but not least, the choice of $\mathbb{A}$ can be inspired by the quantum behavioral decision theoretical literature. In particular, \cite{aerts2009quantum} and \cite{yukalov2010mathematical}, among others,  consider an orthonormal basis $\ket \omega$ for $\omega$ = 0, $\ldots$, $|\Omega|-1$ as states and compute utility of what they call an act represented by the  unitary operator  $\mathbb{A}$ as $\bra \omega \mathbb{A}  \ket \omega$. The operator $\mathbb{A}$ is sometimes called {\it state of mind} by \cite{eichberger2018decision} among others, and takes advantage of the \cite{vonneumann1932mathematische} quantum probability framework.

\medskip

The main motivation is that econometricians worry about model specification errors. In a series of papers Lars Hansen and Thomas Sargent have developed a framework to formalize the links between risk, ambiguity (about Bayesian priors), and misspecification (of the assumed model) and their decision-theoretic, robust control, and statistical foundations (see \cite{hansen2022risk} and references therein).

\medskip

We rely on a framework suggested by \cite{hansen2022risk} who align definitions of statistical
models, uncertainty, and ambiguity with ideas from decision theory that build on representations of subjective and objective uncertainties articulated by \cite{anscombe1963definition}. They adopt a version of \cite{anscombe1963definition} (henceforth AA) where model parameters are {\it states} with $(\Theta, \mathfrak{G})$ a measurable space of potential states. 
Let $\mathscr{P}$ be the set of probability measures over states and for each $p$ $\in$ $\mathscr{P}$, ($\Theta$, $\mathfrak{G}$, $p$) is a probability space.
	
\medskip
	
For each parameter vector $\theta$ $\in$ $\Theta$, we define the vector $\bar{\nu}_{N}(\theta)$ $\in$ $\mathbb{R}^N$ of model-implied price-dividend ratios. Moreover, using the definitions in equation (\ref{eq:finalquadeq}), we have $\mathcal{A}_N(\theta)$	$\bar{\nu}_{N}(\theta)$ =   $b_N(\theta)$ with  $\mathcal{A}_N(\theta)$ = $[I_N - \mathcal{T}_N(\theta)]$ = $[I_N -  \mathcal{H}_N(\theta) \circ \mathcal{M}_N(\theta) \circ \Pi_N(\theta)]$ and therefore $\bar{\nu}_{N}(\theta)$ = $\mathcal{A}_N^{-1}(\theta)b_N(\theta)$. An alternative characterization is: $\mathcal{C}_N(\theta)$ = $\mathcal{B}^{-1}_N(\theta)\mathcal{A}_N(\theta)$ which also encodes all the information about the model for a given parameter vector.

\begin{remark}
	We do not exactly follow the setup of \cite{hansen2022risk}, but instead fashion it to our specific model of interest, namely solving dynamic asset pricing models of the type appearing in equation (\ref{eq:inteq2}). There are some non-trivial deviations, notably in the assumption below, which will be further discussed.
\end{remark}

\medskip

\noindent To avoid additional notation we assume that the price-dividends in $\bar{\nu}_{N}(\theta)$ $\in$ $\mathbb{R}^N$ are sorted from low to high such that we can associate the solution of a model, given $\theta$, with a distribution function over possible {\it prizes} in the \cite{anscombe1963definition} sense. Namely:
\begin{guess}
\label{ass:lowtohigh}
The equilibrium price-dividends in $\bar{\nu}_{N}(\theta)$ $\in$ $\mathbb{R}^N$  represent \cite{anscombe1963definition} prizes. Moreover, for each $\theta$ $\in$ $\Theta$ there is a unique $\bar{\nu}_{N}(\theta)$ and for any two $\theta_1$ and $\theta_2$ $\in$ $\Theta$, $\bar{\nu}_{N}(\theta_1)$ $\neq$ $\bar{\nu}_{N}(\theta_2)$. Hence, the parameter vector relates bijectively to prizes.
	\end{guess}
	
	\medskip
	
\noindent Assumption \ref{ass:lowtohigh} identifies a unique $N$-point vector with each parameter $\theta$. In the next section, we will further explore the decision-theoretic foundations, adopting a quantum computational and informational approach. This will highlight the unique features of QC, and also showcase opportunities that come with it.

\medskip

We conclude with a discussion of multiple models as it will be helpful in the analysis of misspecification risk and ambiguity. We use the notation $N_D$ for dimension of the discrete states and $N_M$ as the number of models. Note that $N_D$ is determined by a quadrature rule. For simplicity we assume that $N_D$ is the same across the $N_M$ different models. 
To address the issue of multiple models we start from a situation where $m$ = 1, $\ldots$, $N_M$ plausible models are put forward and each model $m$ is associated with a measurable space of potential states  $(\Theta_m, \mathfrak{G}_m)$. Let $\mathscr{P}_m$ be the set of probability measures over states and for each $p_m$ $\in$ $\mathscr{P}_m$, ($\Theta_m$, $\mathfrak{G}_m$, $p_m$) is a probability space. This means we have a collection $\{(\mathcal{A}_{N_D}^m(\theta_m), b_{N_D}^m(\theta_m)) : m = 1, \ldots, N_M\}$ or $\{\mathcal{C}_{N_D}^m(\theta_m) : m = 1, \ldots, N_M\}$ (cfr.\ equation (\ref{eq:finalquadequnitnorm})) and associated solutions  $\{\bar{\nu}_{N_D}^m(\theta_m) :  m = 1, \ldots, N_M\}$. 

\medskip

The solution $\bar{\nu}_{N_D}^m(\theta_m)$ has several potential sources of specification error since $\mathcal{A}_{N_D}^m(\theta_m)$  =$[I_N - \mathcal{H}_{N_D}^m(\theta_m) \circ \mathcal{M}_{N_D}^m(\theta_m) \circ \Pi_{N_D}^m(\theta_m)]$. Namely, we may have concerns about $\Pi_{N_D}^m(\theta_m)$ in particular as the persistence of the state process might be in doubt. Similarly, the stochastic discount factor $\mathcal{M}_{N_D}^m(\theta_m)$ could be a source of model error. Lastly, and probably least likely, the payoff function embedded in $\mathcal{H}_{N_D}^m(\theta_m)$ might be prone to mistakes. Note that by implication $b_{N_D}^m(\theta_m)$ is affected by these uncertainties as well.

\begin{remark}
	\label{remark:lotrace}
	\cite{anscombe1963definition} distinguish ``lotteries'' governed by known, i.e.\ objective, probabilities from ``horse races''  with unknown (subjective) probabilities. \cite{hansen2022risk} adopt this framework in the context of statistical inference. We follow a similar approach, though again different and tailored to the specific application we have mind. To that end, we will sometimes split the parameter vector  $\theta$ $\in$ $\Theta$, with $\theta$  $\equiv$ ($\theta_L,\theta_H$) $\in$ $\Theta_L \times \Theta_H$ with the former governed by some objective probability distribution, whereas the latter by a subjective one. Case in point, specifically of interest in the context of our application is  $\mathcal{A}_N(\theta)$ = $[I_N -  \mathcal{H}_N(\theta_L) \circ \mathcal{M}_N(\theta_H) \circ \Pi_N(\theta_L)]$, meaning the econometrician has some objective distribution to assess the model for state dynamics, but has potential doubts about the SDF. Of course, we can also consider the two polar cases $\theta$ $\equiv$ $\theta_L$ and $\theta$  $\equiv$ $\theta_H$. 
\end{remark}

So far we dealt with ambiguity, which is modeled via the measurement operator. What about uncertainty, or in the terminology of \cite{anscombe1963definition}, lotteries? Here we rely on mixed states instead of pure states representing a single asset pricing model. Recall that we considered $N_M$ models: $\bar{\nu}_1(\theta_{1})$, \ldots, $\bar{\nu}_{N_M}(\theta_{N_M})$.  Suppose now that we have some objective probability distribution across these models, attaching  probability $p_j$ to model $j$, with $\sum_j p_j$ = 1. This could come from an asymptotic sampling argument for instance as we will pursue in the next subsection. In quantum mechanics parlance we construct the mixed state, which in AA parlance is a lottery of models $Q$ = $\sum_j p_j P_{\bar{\nu}_j}$, $\sum_j p_j$ = 1 and act on it with	$\mathbb{A}$. Hence, uncertainty is represented by mixed states, with $P_{\bar{\nu}_j}$ the projection operator associated with a particular model $j,$ and ambiguity is formulated through the measurement operator.

\medskip

Inspired by \cite{eichberger2018decision}, who use the same scheme to provide quantum decision-theoretic solutions to the \cite{ellsberg1961risk} paradox, we define quantum ambiguity based on the following superposition of quantum states:
\begin{equation}
	\label{eq:complexdoubts}
	\ket{\mathcal{S}_{d,B}(\alpha,\delta)} =  \alpha \ket{d} + e^{i\delta} \sqrt{1 - \alpha^2} \ket{\bar{\nu}_B}
\end{equation}
and associated acts through the set of self-adjoint projection operators:
\begin{equation}
	\label{eq:complexdoubtP}
	\mathbb{A} = P_{	\ket{\mathcal{S}_{d,B}(\alpha,\delta)}} = 	\ket{\mathcal{S}_{d,B}(\alpha,\delta)} 	\bra{\mathcal{S}_{d,B}(\alpha,\delta)}, \qquad \text{ with } \delta \neq 0.
\end{equation}
which involves two parameters, namely $p$ as in the classical mixture, and $\delta$ pertaining to quantum interference. \cite{eichberger2018decision} emphasize in their analysis the important role to solving Ellsberg's paradox played by the complex-valued interference term $e^{i\delta}$, or put differently, the important role played by $\delta$ in addressing ambiguity. Theorem 4.1 of \cite{ghysels2024measure} shows that:
\begin{eqnarray}
	\langle d - \bar{\nu}(\theta) \vert  P_{\ket{\mathcal{S}_{d,B}(\alpha,\delta)}} \vert d - \bar{\nu}(\theta) \rangle & = &	\text{tr}(P_{\ket{\mathcal{S}_{d,B}(\alpha,\delta)}}P_{d - \bar{\nu}(\theta)})  =   \underset{\text{Classical Ambiguity}}{\underbrace{\alpha^2\text{tr}(P_{d}P_{d - \bar{\nu}(\theta)}) + (1 - \alpha^2) \text{tr}(P_BP_{d - \bar{\nu}(\theta)})}} \notag \\
	& &   + \underset{\text{Quantum Ambiguity}}{\underbrace{(\alpha \sqrt{1 - \alpha^2}) r_{d - \bar{\nu}(\theta)}^d r_{d - \bar{\nu}(\theta)}^B \cos (\delta  + \alpha_{d - \bar{\nu}(\theta)}^{d,B})}} 		\label{eq:theoremabc}
\end{eqnarray}	
where $r_{d - \bar{\nu}(\theta)}^de^{i\alpha_{d - \bar{\nu}(\theta)}^d}$  = $\langle d - \bar{\nu}(\theta) | d \rangle$, $r_{d - \bar{\nu}(\theta)}^Be^{i\alpha_{d - \bar{\nu}(\theta)}^B}$ = $\langle d - \bar{\nu}(\theta) | \bar{\nu}_B \rangle$, and  $\alpha_{d - \bar{\nu}(\theta)}^{d,B}$ = $\alpha_{d - \bar{\nu}(\theta)}^B  - \alpha_{d - \bar{\nu}(\theta)}^d$. 

\setcounter{equation}{0}
\setcounter{table}{0}
\setcounter{figure}{0}
\section{Illustrative Empirical Model Implementation \label{sec:implementation}}

So far we have achieved the following: given a model of the type appearing in equation (\ref{eq:inteq2}), we have a quantum state $\ket{\bar{\nu}(\theta)}$ which encodes the equilibrium asset pricing properties. In experimental physics one builds instruments to {\it measure} the quantum state. When the state of a quantum system is measured, it interacts with a classical measurement device, described mathematically by an operator. For QC applications in economics and finance, we are obviously not in a situation with lab measurement instruments, so we need to think about what to do once we obtain the solution $\ket{\bar{\nu}(\theta)}$. 

\medskip

We will consider five asset pricing models to assess them empirically with the new tools relying on quantum computational solutions. A subsection is devoted to the different models.\footnote{All the codes are posted on Github \url{https://github.com/jackhmorgan/Dynamic_Asset_Pricing_Models}. \label{footnote.codes}}

\subsection{Asset Pricing Models with CRRA and Recursive Utility}

We start with equation (2) of \cite{hansen2008consumption} applied to quarterly log dividend growth.\footnote{\cite{hansen2008consumption} focus on consumption growth and examine the risk pricing of future cash flows. For our purpose, it will be convenient to start with an equation linking dividend growth directly to a state variable process.} Namely let $d_t$ = $\log D_t$, and:
\begin{eqnarray}
	\label{eq:logdivgrowth}
	d_{t+1} - d_t &=& a+x_{t+1}+be_{t+1}\\
	x_{t+1} &=& \rho_1 x_{t}+c \epsilon_{t+1}  \notag
\end{eqnarray}
We collect the parameters of the above model into the vector $\theta_L$ = $(a, b, c, \rho_1)^\prime$.
Assuming standard normal errors, we can estimate the parameters of the model (\ref{eq:logdivgrowth}) via Maximum Likelihood.\footnote{Details about the estimation appear in Online Appendix Section \ref{appsec:DAPM}. The parameter estimates appear in Table \ref{tab:Solutions_div}.} To implement the various measurement operators, including those pertaining to model uncertainty appearing in Section \ref{subsec:measurementambiguity}, we rely on the asymptotic distribution theory for MLE to characterize model uncertainty about equation (\ref{eq:logdivgrowth}). Namely, we assume that the model is correctly specified, but we are uncertain about its parameter values. Hence, we don't question neither the functional form nor the distributional assumptions. This is done for the sake of simplicity and it reflects the notion of AA lotteries.  This will create mixed quantum states, and therefore reflect uncertainty.  Namely, we  create one thousand draws of parameter models $\{\theta_i, i = 1, \ldots 1000 \}$ and construct a 1000 models for log dividend growth. More specifically , we draw one thousand $\theta_L^i \sim N(\hat{\theta}_L, \hat{\Sigma})$,  excluding draws with $\rho_1 < 0$ and $\rho_1 > 1$. The collection of models therefore represents statistical uncertainty, or mixed states, since we assume the model specification is correct and asymptotic distribution theory provides us guidance about parameter uncertainty.\footnote{We can characterize and visualize the uncertainty using the
Kullback-Leibler divergence $KL_i=\frac{1}{2}[(\theta_L^i - \hat{\theta}_L)' \hat{\Sigma}^{-1}(\theta_L^i - \hat{\theta}_L))]$. The fitted distribution appears in Online Appendix Figure \ref{graph:KL_dist}. This statistical ensemble is the basis to construct a mixed state benchmark model.}

\medskip

The log of the one period stochastic discount factor $s_{t+1,t}$, again following the setup of \cite{hansen2008consumption}, is as follows (with parameter estimates discussed in Online Appendix Section  \ref{appsec:DAPM}):
\begin{equation}
	\label{eq:logSDF}
	s_{t+1,t} = -0.8974 + 1.2038 x_t + \xi w_{t+1}.
\end{equation}
We focus only on the parameter $\xi$, and will consider two cases, determined by assumptions regarding the preferences of the representative agent. Namely, under CRRA preferences $\xi$ = $-0.03630\gamma$, using the parameter estimates reported in Online Appendix Table \ref{tab:Solutions_div}, with $\gamma$ the coefficient of relative risk aversion and the slope is the estimate of the parameter pertaining to the dividend shock in equation (\ref{eq:logdivgrowth}). The second case involves recursive utility with the intertemporal elasticity of substitution  equal to one, where  $\xi$ =  $0.0412 - 0.0775\gamma$, using parameter estimates reported in  Online Appendix Section  \ref{appsec:DAPM}.  

\medskip

In Appendix Section \ref{appsec:DAPM} we also describe the computations for a single model. All calculations use a four by four transition matrix, $N$ = 4, which results in a 16-dimensional quantum linear system problem to create the state  $\ket{d - \bar{\nu}(\theta)}.$  We classically prepare the quantum state vector and calculate the expectation value of the observable using $Qiskit.$\footnote{Qiskit is an open-source software development kit for programming quantum computers. It provides tools to create, manipulate, and run quantum programs on IBM quantum devices. It is also widely used on other quantum hardware platforms and is the leading Python software development kit for open source quantum development.} As noted earlier, our classically computed expectation value is equivalent to the result of a hypothetical HHL implementation without hardware noise nor algorithmic error. Such an implementation would require a large number of evaluation qubits for QPE and a noiseless processor. Needless to say, our classical calculations do not pose an exponential speedup. However, they serve as a proof of concept for quantum implementations that will provide an exponential speedup once hardware is up to the task. 

\medskip

We fix $\gamma$ = 10 and define the benchmark model as the CRRA specification with $\ket{\bar{\nu}_B}$ = $\sum_j p_j \ket{d - \bar{\nu}_j}$ using the mixed state weights obtained from the Kullback-Leibler divergence, the weights are $p_j$ = $KL_j/\sum_j KL_j$. A few words regarding doubt about the data quantum state $\ket d$. Recall that the functional form in equation (\ref{eq:logdivgrowth}) is given and the parameters are uncertain. Nevertheless, one may still argue that the dividend growth data is a source of concern and entertain the thought that observed dividends are only a noisy proxy for economy-wide asset payoffs. This concern may be embedded in the measurement operator as doubt about the data.

\medskip

Using the benchmark model we evaluate the IES models with $\gamma$ = 2, 5 and 10 as target models. Hence, in our empirical illustration we have $\ket{d}$ and  $\ket{\bar{\nu}_B}$ the aforementioned CRRA model with $\gamma$ = 10 and the mixed state statistical ensemble pertaining to dividend parameter uncertainty. Finally, the targets are the pricing errors $\ket{d - \bar{\nu}(\theta)}$ for the three long-run risk models with the same mixed states representing the parameter uncertainty. 

\medskip

\begin{figure}
	\begin{center}
		\caption{Illustrative Empirical Models: CRRA versus Recursive Preferences	\label{fig:classvsquantEMP}} 
		\begin{subfigure}{0.42\textwidth} 
			\includegraphics[width=\textwidth]{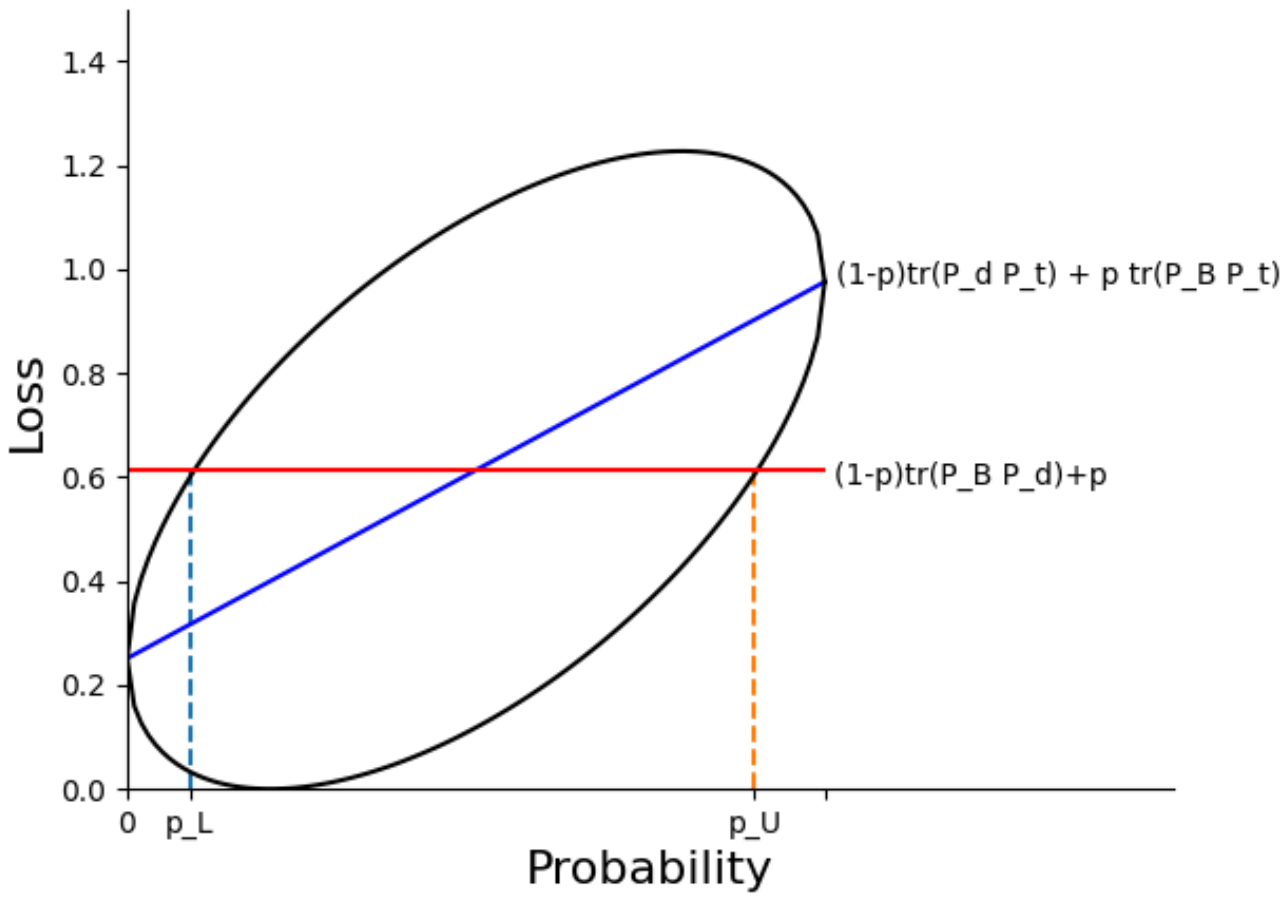}
			\caption{Benchmark CRRA with p = 0.5  \\ \qquad  \qquad Target IES = 1, $\gamma$ = 2 \\
				\qquad $p_L$ = 0.09, $p_C$ = 0.49, $p_U$ =0.90} 
		\end{subfigure}
		\begin{subfigure}{0.42\textwidth} 
			\includegraphics[width=\textwidth]{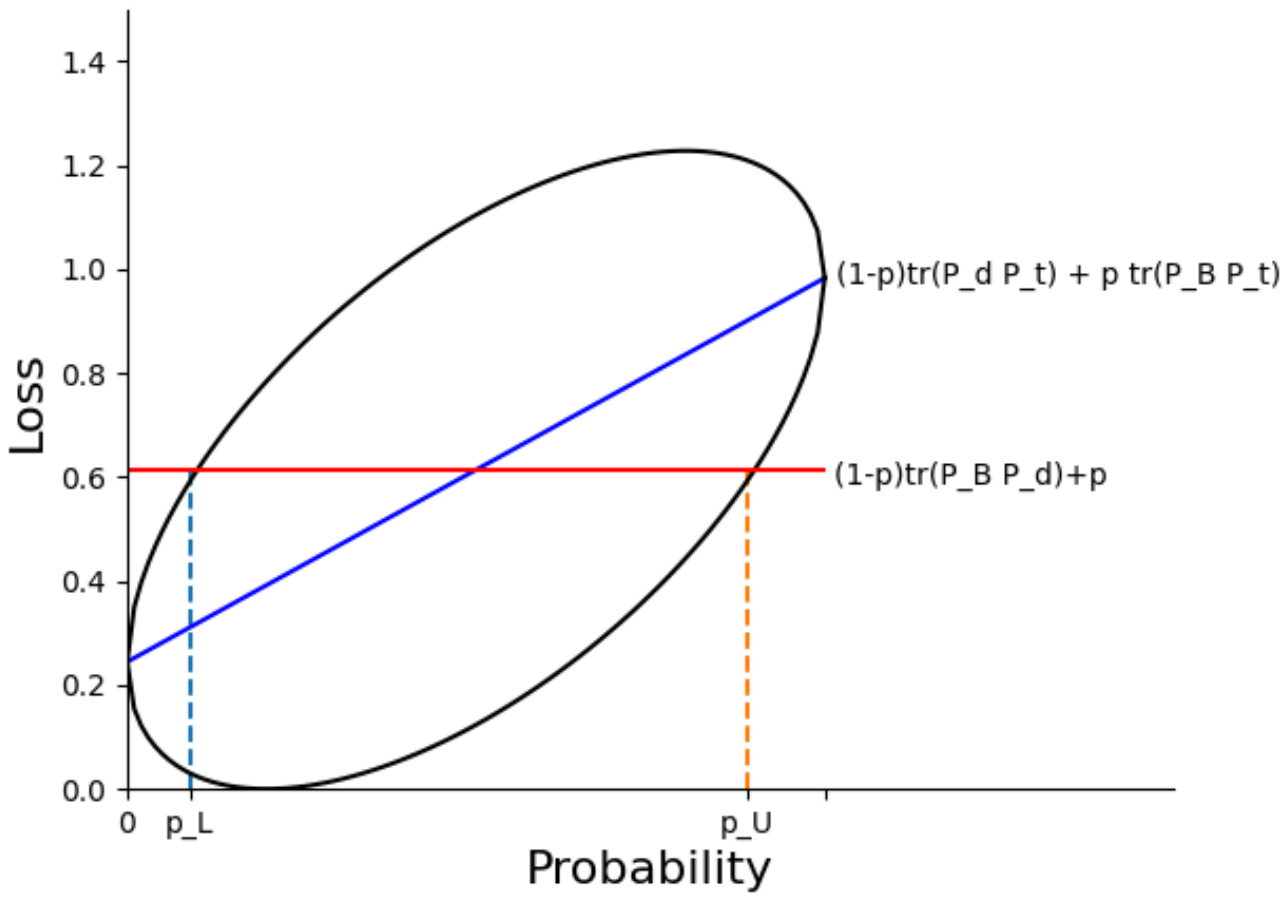}
			\caption{Benchmark CRRA with p = 0.5 \\ \qquad  \qquad Target IES = 1, $\gamma$ = 10
				\\
				\qquad $p_L$ = 0.09, $p_C$ = 0.49, $p_U$ =0.89} 
		\end{subfigure}
		\begin{subfigure}{0.42\textwidth} 
			\includegraphics[width=\textwidth]{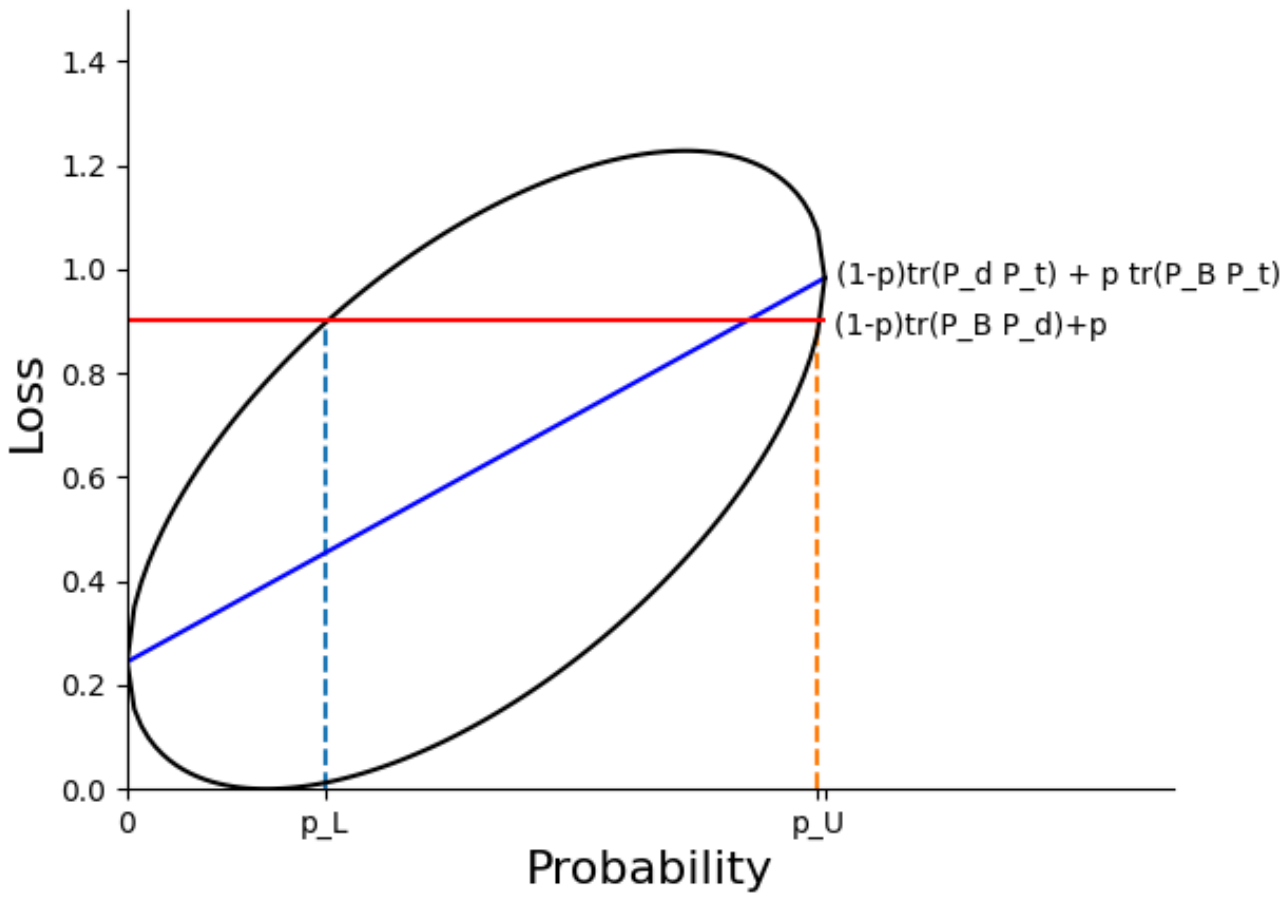}
			\caption{Benchmark CRRA with p = 0.9  \\ \qquad  \qquad Target IES = 1, $\gamma$ = 10\\
				\qquad $p_L$ = 0.28, $p_C$ = 0.89, $p_U$ =0.99} 
		\end{subfigure}	
		\begin{subfigure}{0.42\textwidth} 
			\includegraphics[width=\textwidth]{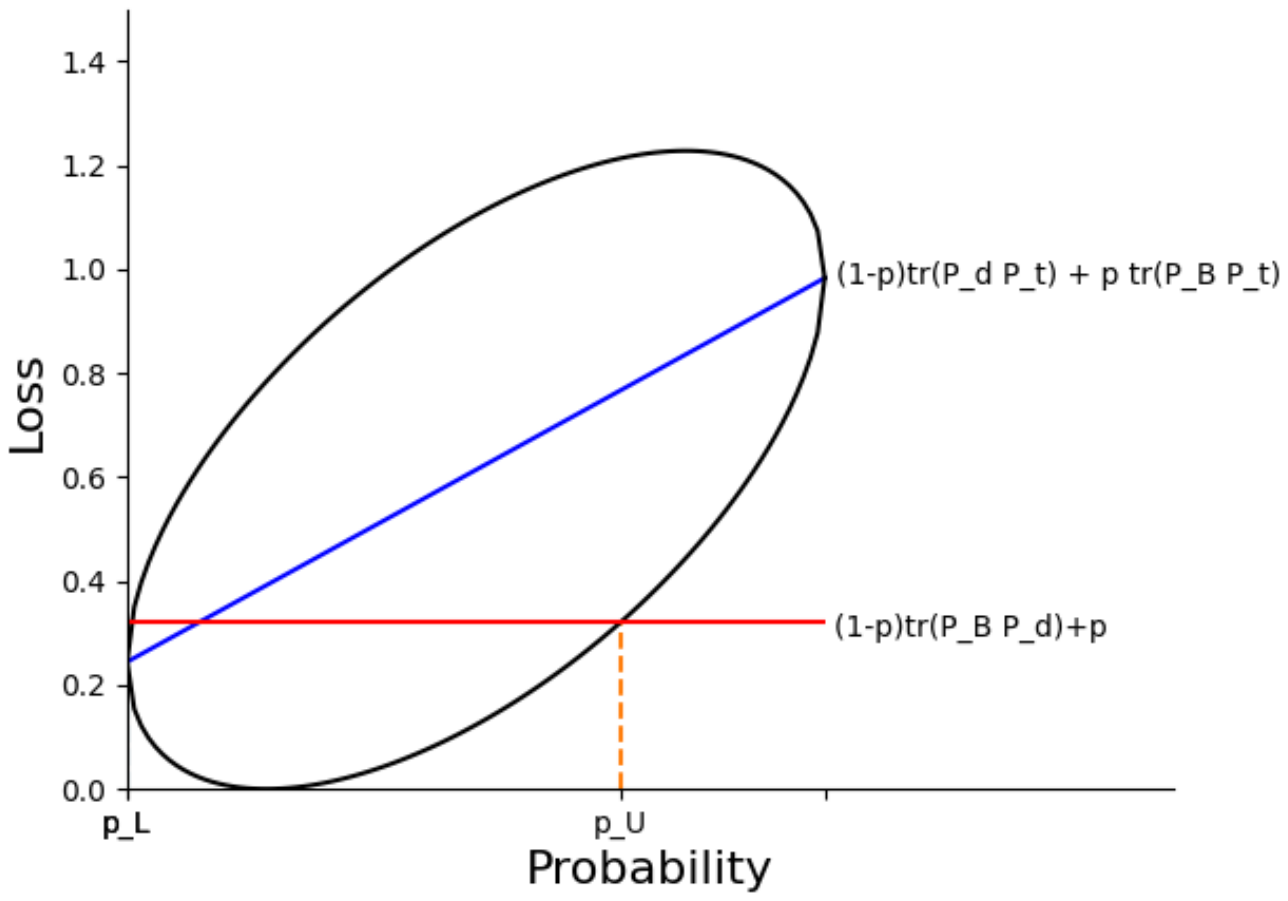}
			\caption{Benchmark CRRA with p = 0.1 \\ \qquad  \qquad Target IES = 1, $\gamma$ = 10 \\
				\qquad $p_L$ = 0.00, $p_C$ = 0.10, $p_U$ =0.71} 
		\end{subfigure}	
	\end{center}
	\begin{center}
		\begin{subfigure}{0.43\textwidth} 
			\includegraphics[width=\textwidth]{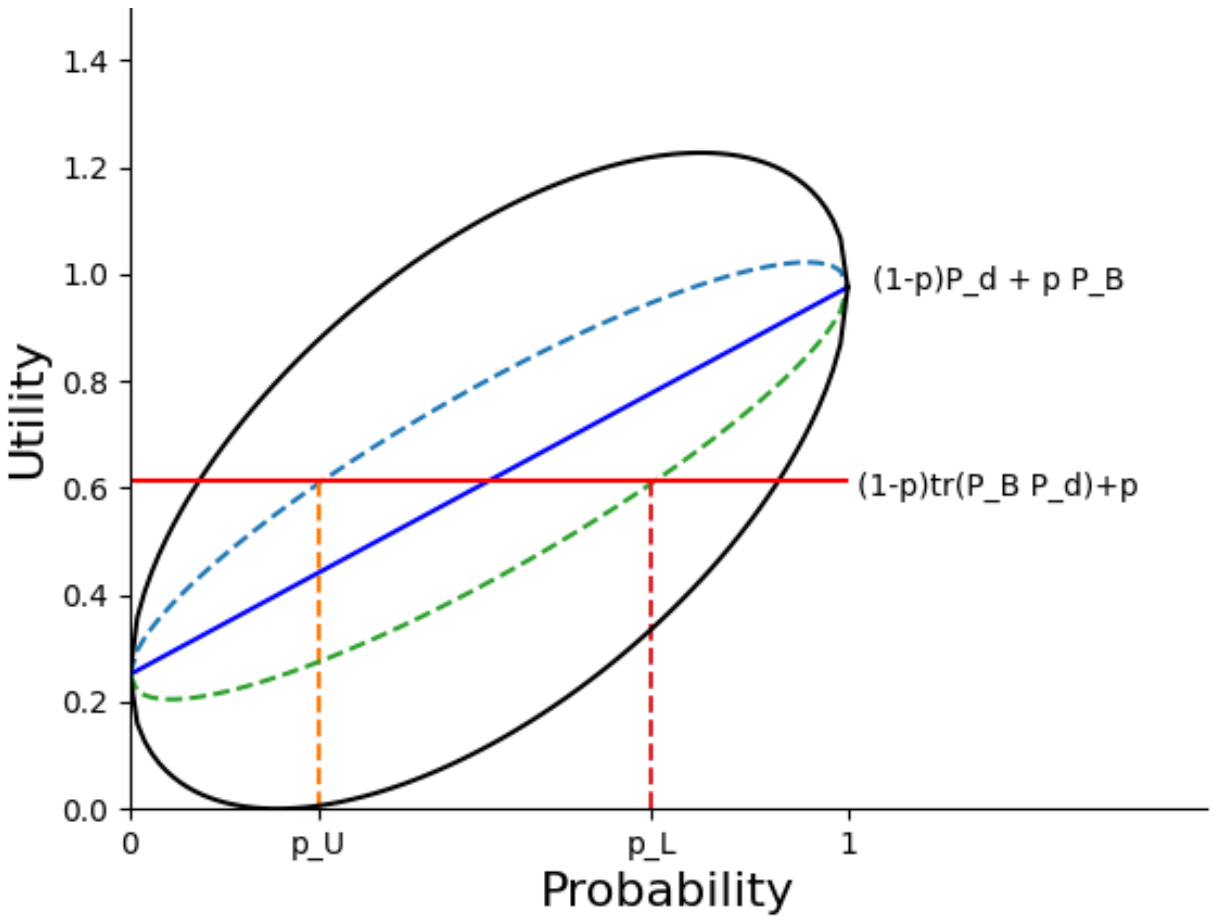}
			\caption{Changing $P_{\ket{\mathcal{S}_{d,B}(\alpha,\delta)}}$ parameters \\ \quad Benchmark/Target as in Panel (b) \\ \qquad \quad $p_L$ = 0.26 \, $p_U$ = 0.72} 
		\end{subfigure}	
	\end{center}
\end{figure}

We turn our attention to Figure \ref{fig:classvsquantEMP} where we consider decisions about model selection. The reference horizontal line corresponds to 0.5tr$\left( P_d P_B\right)$ + 0.5 tr$\left(P_B P_B \right)$ = 0.5 + 0.5tr$\left( P_d P_B\right).$ Hence, we take a 50-50 view of the benchmark model $P_B$ versus the data $P_d.$ The upward sloping line represents the target model pricing error against the classical mixture measurement operator defined by equation 
(\ref{eq:complexdoubts}).
In panels (a) and (b) of Figure \ref{fig:classvsquantEMP} we display two target models, both with EIS = 1 and $\gamma$ = 2 versus $\gamma$ = 10. We start with the classical ambiguity measurement operator with a mixture $p.$
Above the point $p_C$ we opt for the benchmark model which has a lower expected loss and left of $p_C$ we opt for the target model. Hence with a classical measurement operator to look at the target model with a lot doubts about the benchmark model, i.e.\ $p$ close to zero, we opt for the target model. Conversely, with strong doubts about the data we stick to the benchmark model in this case. The intersection point $p_C$ = 0.49. Not much different from the 50-50 mix.

\medskip

Quantum ambiguity provides a more interesting and rich model selection decision process. The oval-shaped plot represents the extremes of the ambiguity component $(\alpha \sqrt{1 - \alpha^2}) r_{d - \bar{\nu}(\theta)}^d r_{d - \bar{\nu}(\theta)}^B \cos (\delta  + \alpha_{d - \bar{\nu}(\theta)}^{d,B}).$ Those extremes are obtained for $\cos (\delta  + \alpha_{d - \bar{\nu}(\theta)}^{d,B})$ = 1 (and therefore $(\delta  + \alpha_{d - \bar{\nu}(\theta)}^{d,B})$ = 0) and $\cos (\delta  + \alpha_{d - \bar{\nu}(\theta)}^{d,B})$ = -1 (and therefore $(\delta  + \alpha_{d - \bar{\nu}(\theta)}^{d,B})$ = $\pi$). Therefore the oval shape height is determined by the range is $[-(\alpha \sqrt{1 - \alpha^2}) r_{d - \bar{\nu}(\theta)}^d r_{d - \bar{\nu}(\theta)}^B,(\alpha \sqrt{1 - \alpha^2}) r_{d - \bar{\nu}(\theta)}^d r_{d - \bar{\nu}(\theta)}^B]$. First, we call the interval $[p_L, p_U]$ as inconclusive. By that we mean that unless we pick specific parameters $p$ and $\delta,$ we don't have a quantum unambiguous answer between keeping the benchmark model or the new target model when measuring the quantum state of the latter with a quantum ambiguous measurement operator appearing in equation (\ref{eq:complexdoubts}) involving the data and benchmark model. We observe that the length of the interval between $p_L$ and $p_U$ is unchanged as we increase the parameter $\gamma$ in the SDF. This result is not surprising given what we know about the challenges pertaining to the estimation of the SDF. Increasing $\gamma$ in this case does not move the needle so to speak. Note that $P_L$ = 0.09, so that the target model is adopted if we have at least 91 percent confidence in the data. Conversely, we stick to strong beliefs about the benchmark model in excess of 90 percent and therefore very little confidence in the data, we quantum unambiguously disregard the target model.

\medskip

Panels (c) and (d) modify the classical mix of data and benchmark as a reference point. In panel (c) we begin with little confidence in the data $p$ = 0.9. Against that reference, we won't settle quantum unambiguously for the benchmark model when measuring the evidence of the target model since $p_U$ = 0.99. In panel (d) we change to a reference will strong (classical) confidence in the data. Unless, we take a strong view in favor of the benchmark model $p_U$ = 0.71, we go for the target model no matter what $p$ we pick, since $p_L$ = 0.

\medskip

Of course, we can play with $p$ and $\delta$ as we proceed to measurement of the target model quantum state and make model decisions guided by expected loss. Recall that the oval-shaped plot represents the extremes of the ambiguity component  $[-(\alpha \sqrt{1 - \alpha^2}) r_{d - \bar{\nu}(\theta)}^d r_{d - \bar{\nu}(\theta)}^B,(\alpha \sqrt{1 - \alpha^2}) r_{d - \bar{\nu}(\theta)}^d r_{d - \bar{\nu}(\theta)}^B]$. Panel (e) of Figure \ref{fig:classvsquantEMP} shows a reduced oval-shaped region of quantum ambiguity and by implication a smaller inconclusive interval. Note that  $r_{d - \bar{\nu}(\theta)}^d,$ $r_{d - \bar{\nu}(\theta)}^B$ and  $\alpha_{d - \bar{\nu}(\theta)}^{d,B}$ = $\alpha_{d - \bar{\nu}(\theta)}^B  - \alpha_{d - \bar{\nu}(\theta)}^d$ are determined by the choice of the benchmark model, the target model and the features of the data.  For example, the econometrician/decision maker might commit to a narrower notion of quantum ambiguity and consider a small range for $\delta.$ In Panel (e), with the same target and benchmark as in Panel (b), we restricted $\delta$ to vary between $3 \pi/8$ and $5 \pi/8.$ .  Now, we have $p_L$ = 0.26 and $p_U$ = 0.72 instead of $p_L$ = 0.09 and $p_U$ =0.89 in Panel (b).  Another equally important input is the role of  $\alpha_{d - \bar{\nu}(\theta)}^{d,B}$ = $\alpha_{d - \bar{\nu}(\theta)}^B  - \alpha_{d - \bar{\nu}(\theta)}^d,$ i.e.\ the angle between the data and the benchmark in this example versus the target model and the benchmark model. Therefore, changing the benchmark and/or target model change the shape of the ambiguity area. The inputs are relatively straightforward however: (a) the data, (b) the target model, (c) the benchmark model, and (d) two parameters $p$ and $\delta.$

\subsection{Adding Stochastic Volatility \label{subsec:SV}}

\cite{bansal2004risks} introduced a prominent asset pricing model that helps explain various puzzles, including the equity premium puzzle and risk-free rate puzzle, by introducing long-run consumption risks and time-varying volatility. So far we dealt with the Epstein-Zin type recursive preferences but did not have a data generating process with stochastic volatility (SV). We now add SV to our analysis. In the original \cite{bansal2004risks} model the expected growth rate of (log consumption - which we replace by log dividends) evolves over time as follows: $d_{t+1}$ -  $d_{t}$ = $a$ + $x_t$ + $\sigma_{t} e_{t+1},$ where $x_t$ is a persistent process as in equation (\ref{eq:logdivgrowth}) and $\sigma_{t}$ represents stochastic volatility with the following dynamics:
$\sigma^2_{t+1}$ = $\sigma^2$ + $\nu (\sigma^2_t - \sigma^2)$ + $\sigma_w \epsilon_{t+1}.$

\medskip

The challenge is how to translate the SV setup in \cite{bansal2004risks} to one amenable to a quantum circuit formulation. To do so, we adopt the approach of \cite{ghysels2023quantum} to Markov regime switching models and \cite{ghysels2025quantum} to the MLE of SV models. Key to translating the continuous state SV model to a discrete state Markov process. For our purpose can think of a two state model, for convenience called the good economy $G$ and the other the bad economy $B$ state with transition matrix:
\begin{equation}
	\label{eq:transition}
	\mathscr{T} = \left[
	\begin{array}{cc}
		1 - p_{GB} & p_{GB} \\
		p_{BG} & 1 - p_{BG}
	\end{array}
	\right],
\end{equation}
where $1 - p_{GB}$ is the probability to stay in the good state,  as $p_{GB}$ is the probability to move to the the bad state. Likewise, we have 1 - $p_{BG}$ as the probability to stay in the bad state and $p_{BG}$ is the probability to move to the good state. In our computations we will only look at the implied ergodic distribution: $[\pi_G, 1 - \pi_G]$ where $\pi_G$ = $p_{BG}/(p_{BG} + p_{GB}).$ The advantage of only selecting $\pi_G,$ instead of the full transition matrix, is that we have one less parameter to fix in the computations.

\medskip

Next, we need to define the DGP for the good and bad economies. This warrants a discussion of the parameters for the dividend growth process and their ML point estimates. Since we assume that the SV model is independent of $x_t$ we can assume that the ML estimates are consistent, but not asymptotically efficient. Hence, while we can rely on a quasi-MLE argument to use the parameters  $\theta_L$ = $(a, b, c, \rho_1)^\prime,$ but the use of the MLE asymptotic distribution to characterize model uncertainty about equation (\ref{eq:logdivgrowth}) in principle needs a QMLE modification as the outer-product of scores and Hessian differ. However, since our focus is not inference, we use keep the MLE distribution to facilitate the comparison with models without SV - maintaining sources of uncertainty fixed. The QMLE interpretation also allows us to anchor one of the two regime volatilities since: $b$ = $\pi_G \times b_G$ + $(1 - \pi_G) \times b_B,$ with $b_B$ $>$ $b$ $>$ $b_G,$ the latter being the volatility associated with the good economy. We normalize $b_G$ to a fraction $\gamma_G \times b,$ with 0 $<$ $\gamma_G$ $<$ 1, such that 
$b_B$ = $b(1 - \gamma_G \times \pi_G)/(1 - \pi_G).$ As a result, we only need to fix $\pi_G$ and $\gamma_G$ to study models with SV. Further details appear in Online Appendix Section \ref{appsubsec:SV}.

\begin{figure}
	\begin{center}
		\caption{Classical Ambiguity Comparison: Constant vs. Stochastic Volatility.	\label{fig:classvsquantSV}} 
		\begin{subfigure}{0.42\textwidth} 
			\includegraphics[width=\textwidth]{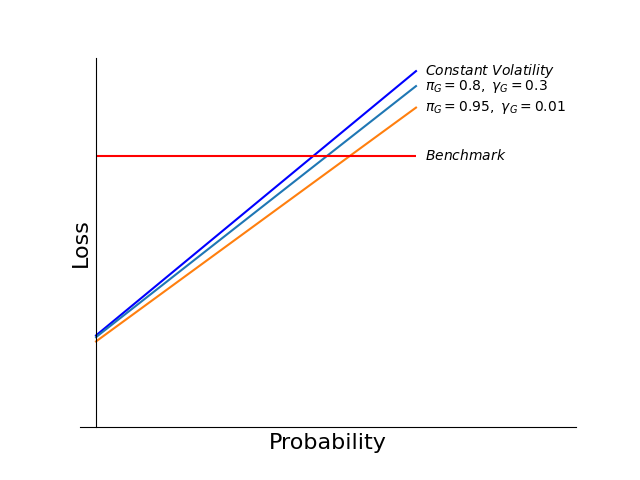}
			\caption{Comparison between the model from Figure \ref{fig:classvsquantEMP} and the models in panels (a) and (b) from this figure.} 
		\end{subfigure}	
	\end{center}
\end{figure}

We compare the classical ambiguity of constant and stochastic volatility models in Figure \ref{fig:classvsquantSV}. As in Figure \ref{fig:classvsquantEMP}, we use a benchmark model with CRRA utility and $\gamma$=10. The three featured models all use IES = 1 and $\gamma$ = 2. We begin with the constant volatility model featured for reference. The two stochastic volatility models use $\pi_G = 0.8$, $\gamma_G = 0.3$ and $\pi_G = 0.95$, $\gamma_G = 0.01$ respectively. A constant volatility model is identical to a stochastic volatility model with $\gamma_G$ = 1. Thus, it is not surprising that the loss from the $\gamma_G$ = 0.3 model is closer to the constant volatility model than that of the $\gamma_G$ = 0.01 model. Around each classical ambiguity line exists a quantum ambiguity ellipse which is similar to the bounds shown in Figure \ref{fig:classvsquantEMP}, however we omit these bounds for the stochastic volatility models for visual clarity.

\medskip

Most importantly, however, is the fact that adding stochastic volatility hardly moves the needle in terms of model fit as shown in Figure  \ref{fig:classvsquantSV}. Indeed, the models are quite similar to the constant volatility specification.

\bigskip

\subsection{Asset Pricing with Rare Disasters \label{subsec:raredis}}

Rare disaster models have been considered by \cite{rietz1988equity}, \cite{barro2006rare}, \cite{gourio2012disaster}, among others. They feature appealing theoretical and empirical asset pricing properties.
Before studying such models, we appraise the models considered so far using the tail event operators introduced in Section  \ref{subsec:benchmarks}. In particular, we use $\mathbb{A}$ = $\ket{u_0} \bra{u_0}$ = $P_0,$ and compute the expectation value, using equation (\ref{eq:utility}):
\begin{equation}
	\label{eq:utilitytail}
	\langle d - \bar{\nu}(\theta) \vert \mathbb{A}\vert d - \bar{\nu}(\theta) \rangle = \text{tr}(P_0 P_{d - \bar{\nu}(\theta)}) \quad \text{ with } P_{d - \bar{\nu}} = \vert d - \bar{\nu}(\theta) \rangle \langle d - \bar{\nu}(\theta) \vert, P_0 = \ket{u_0} \bra{u_0}.
\end{equation}
The results appear in Table \ref{tab:tailexp}. The results show a vast improvement in terms of expectation values when looking at tail event measurement operators applied to the rare disaster model we are about to introduce.
\begin{table}[h]
	\centering
	\caption{Tail Event Measurement Expectation Values}
	\label{tab:tailexp}
	\begin{tabular}{l c c c}
		Models & Without SV & $\pi_G$=0.8 & $\pi_G$ = 0.95 \\  
		 &  & $\gamma_G$ = 0.3 & $\gamma_G$ = 0.01 \\  
		\hline 
		CRRA & 0.38 & 0.29 & 0.23 \\  
		IES = 1, $\gamma$ = 2 & 0.37 & 0.31 & 0.24 \\  
		IES - 1, $\gamma$ = 10 & 0.43 & 0.32 & 0.28 \\  
		Rare disasters & 0.03 & - & - \\
	\end{tabular}
\end{table}
In particular, we focus on the model with variable rare disasters proposed by \cite{gabaix2012variable}. It is a model where despite the underlying
stochastic process being nonlinear and non-Gaussian, one has closed-form solutions. 

The log of the stochastic discount factor, using again log dividend growth instead of consumption as in the previous section, is:
\begin{equation}
	\label{eq:aprare1}
	s_{t+1} -\rho -\gamma(d_{t+1} - d_t) = -\rho -\gamma a = \left\{
	\begin{array}{c}
		0\text{ if no disaster  \ } \\
		-\gamma \log B_{t+1} \text{ if disaster \ }%
	\end{array}
	\right.
\end{equation}%
where $a$ is the average log dividend growth as appearing in Online Appendix equation (\ref{appeq:logdivgrowth}). \cite{gabaix2012variable} defines a process $H_t$ called resilience with the following dynamics:
\begin{equation}
	\label{eq:apprare2}
\hat{H}_{t+1} = \frac{1+H_*}{1+H_t} \exp{(-\phi_H)} \hat{H}_t + \varepsilon_{t+1}^H
\end{equation}
where $H_t$ = $H_*$ + $\hat{H}_t.$ In the above equation $H_*$ is a constant, with $\phi_H$ the speed of mean reversion at $H_t$ = $H_*.$  The equilibrium price–dividend ratio at time $t$ depends
only on $\hat{H}_t$ and is independent of the distribution of $\varepsilon^H_t,$ (see equation (13) in \cite{gabaix2012variable}). Assuming the probability of disaster $p_t$ = $p$  and $B_t$ = $B$ for all $t,$ one has $H_*$ = $p(B^{1- \gamma} - 1).$ In Online Appendix Section \ref{appsec:raredis} we provide the details regarding the numerical implementation of above model. 

\begin{figure}
	\begin{center}
		\caption{Rare Disasters Model	\label{fig:classvsquantRD}} 
		\begin{subfigure}{0.42\textwidth} 
			\includegraphics[width=\textwidth]{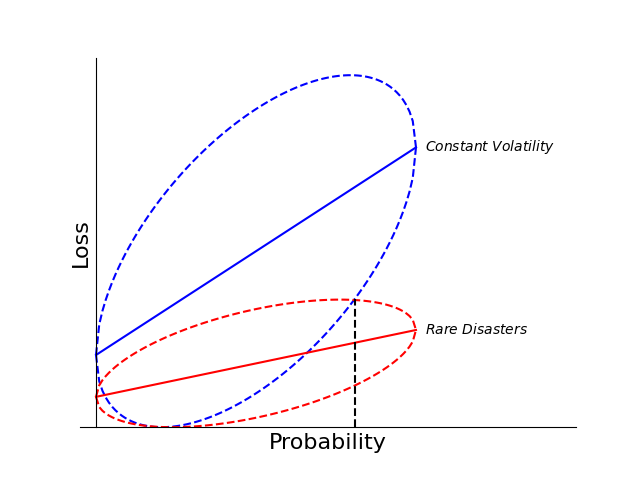}
			\caption{Benchmark CRRA with $\gamma$=10, \\ Target Constant Volatility CRRA with $\gamma$ = 2, \\ Target Rare Disasters model \\
			Intersecting at p = 0.81} 
		\end{subfigure}	
		\begin{subfigure}{0.42\textwidth} 
			\includegraphics[width=\textwidth]{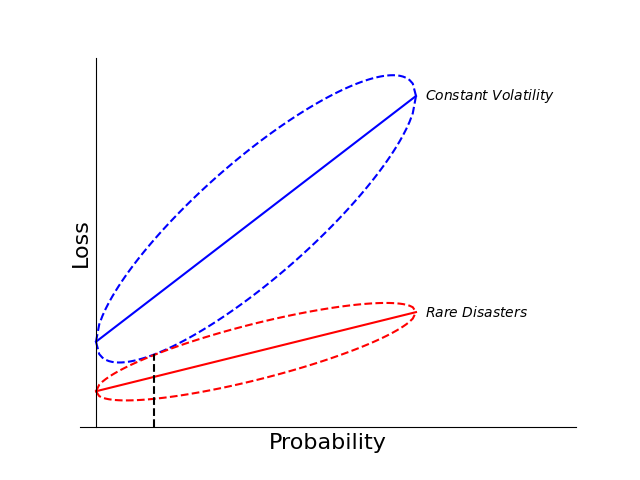}
			\caption{Benchmark CRRA with $\gamma$=10, \\ Target Constant Volatility CRRA with $\gamma$ = 2, \\ Target Rare Disasters model \\
			Quantum uncertainty $\delta$ of $\pi/2$ $\pm$ $0.46$ }
		\end{subfigure}	
	\end{center}
\end{figure}

In Figure \ref{fig:classvsquantRD} we show the same classical and quantum ambiguity analysis for the model with rare disasters. In Panel (a) we look at the maximal ambiguity contour and we see that only for $p$ = 0.81 the rare disaster model is unambiguously better than the Epstein-Zin type preference specification. However, if we entertain a more restrained attitude towards ambiguity, as depicted in Panel (b) where $\delta$ takes values in the interval $\pi/2$ $\pm$ $0.46,$ then we find that the rare disaster model is unambiguously better than the asset pricing model based on recursive utility stochastic discount factors. This plot is consistent with the tail event results in Table \ref{tab:tailexp} which suggest that the rare disasters model outperforms the constant and stochastic volatility examples. There is one caveat about this finding, namely as explained in Online Appendix Section \ref{appsec:hhl-feasibility},  the rare disasters model features a matrix that is less suitable (in terms of condition number and sparsity) for HHL. 
The findings in Figure  \ref{fig:classvsquantRD} are striking and show the appeal of the model evaluations tools put forward in our paper.

\section{Conclusions and Future Research \label{sec:concl}}

We took advantage of a paradigm shift in computer science to advance a new approach to solving dynamic asset pricing models. Along the way we faced the challenge of dealing with quantum state output and suggested a new measurement approach based on model selection criteria. The latter are based on notions of quantum ambiguity that are novel to the literature.

\medskip

Missing from our analysis is the estimation of parameters prior to model selection. Unfortunately, there is no fully developed theory yet to address this. In principle there is the potential of asymptotic expansions along three dimensions. Two are standard in the econometrics literature, while the third is not - at least at first sight. First a few words about the two familiar ones. There is $N$, pertaining to the quadrature-based discretization studied by \cite{tauchen1991quadrature} with its approximation error diminishing, under suitable regularity conditions, as $N$ $\rightarrow$ $\infty$. Unfortunately, making  the step to quantum computing means infinite dimensional Hilbert spaces which pose their own challenges. While finite-dimensional and infinite-dimensional Hilbert spaces share some similarities, such as the inner product structure, they also exhibit crucial differences in terms of completeness, basis (e.g.\ countably infinite), among others. It is beyond the purpose of this paper to explore such extensions. Next is the asymptotics pertaining to the data, in our case underpinning the quantum state $\ket d$. A classic textbook discussion regarding asymptotics appears for instance in \cite{silverman2018density}.

\medskip

This bring us to the asymptotic expansion along the third and final dimension: $S$ the number of ``shots'' or repetitions of the quantum circuit to compute the expectation values.\footnote{There are some similarities with the econometrics literature on simulation-based estimation, see e.g.\ \cite{duffie1993simulated}.} There is a small, but growing, literature on so called quantum statistics. \cite{helstrom1969quantum} and \cite{holevo1982probabilistic} cover the basic foundations, but as \cite{gill2001asymptotics} put it: {\it  Quantum statistics mainly consists of exact results in various rather special models} (although written in 2001, it is still largely true). There are some results on MLE of a parameter say $\theta$ determining a quantum state $\ket{\bar{\nu}(\theta)}$ - for a correctly specified model - and quantum Cram\'er-Rao lower bounds as $S$ $\rightarrow$ $\infty$, see e.g.\  \cite{demkowicz2020multi} for a recent survey. However, much work remains to be done connecting the decision-theoretic foundations of quantum measurement presented in this paper with quantum statistical analysis.

\section{Data Availability Statement}
The data that support the findings of this study are openly available in a public repository at \url{https://github.com/jackhmorgan/Dynamic_Asset_Pricing_Models}.

\newpage

{\small

}

\appendix

\section{Dynamic Asset Pricing Models - Technical Details \label{appsec:DAPM}}

\subsection{A Review of Quadrature Methods\label{appsubsec:quad}}

We start with a probability space, namely, a triple ($\Omega, \mathcal{F}, \mathcal{P}$) where $\Omega$ is a collection of discrete time ($t$ $\in$ $\mathbb{N}$) infinite sequences in $\mathbb{R}^{n_y}$, $\mathcal{F}$ is the smallest sigma algebra of events in $\Omega$ and $\mathcal{P}$
assigns probabilities to
events. The state of the economy is described by the $n_y$-dimensional stationary stochastic process $\{y_t \in \mathbb{R}^{n_y} : t = 1, \ldots, \infty\}$. Also of interest is the sigma filtration $\mathcal{F}_t$ and associated $\mathcal{P}_t$ pertaining to events up to time $t$.
We focus on a single asset with ex-dividend price $p_{t}$ at time time $t$  and future dividend stream $\{d_{t+k}, k = 1, \ldots, \infty\}$. Denote the price-dividend ratio by $\nu_{t}$ = $p_{t}/d_{t}$. Given a representative agent's time $t$ stochastic discount factor (SDF) $m(y_{t+1}, y_t)$, we have:
\begin{equation}
	\label{appeq:mrs}
	\nu_{t} = \mathbb{E}_t \left[ (1 + \nu_{t+1}) h_{t+1} m( y_{t+1}, y_t)\right]
\end{equation}
where $h_{t+1}$ = $d_{t+1}/d_{t}$ = $h(y_{t+1})$ is the dividend growth and $\mathbb{E}_t[\cdot]$ the conditional expectation given the $\mathcal{F}_t$ filtration. We restrict our attention to stationary Markov processes with conditional distribution of $y_t$ given its  past $x_{t-1}$ = $\{y_{t-1}\}$  given by $f(y_t|x_{t-1})$. We can write equation (\ref{eq:mrs}) of the paper in integral form as follows:
\begin{eqnarray}
	\label{appeq:mrsint}
	\nu(x) & = & \int [1 + \nu(y)] \psi(y,x) f(y|x) dy \notag \\
	& = & \int [1 + \nu(y)] \psi(y,x) \frac{f(y|x)}{\omega(y)} \omega(y) dy \qquad \text{ for some } \omega(y) > 0 \notag \\
	\underset{\nu}{\underbrace{\nu(x)}}  & = & \underset{\mathcal{T}[\nu]}{\underbrace{\int \nu(y) \psi(y,x) \frac{f(y|x)}{\omega(y)} \omega(y) dy}} + \underset{g}{\underbrace{\int  \psi(y,x) \frac{f(y|x)}{\omega(y)} \omega(y) 	 dy }}
\end{eqnarray}
where $\nu(x):$ $\mathbb{R}^{n_y}$ $\rightarrow$ $\mathbb{R}$ is the price-dividend ratio as a function of the current state, while $\psi(y,x)$ = $h(y)$ $\times$ $m(y,x)$ with $h(y)$ the dividend growth as function of the future state and $m(y,x)$  the SDF. 
\begin{remark}
	Estimation and inference of the asset pricing model appearing in equation (\ref{appeq:mrsint}) require a parametric specification for the functions  $\psi(y,x)$,  $f(y|x)$, and $\omega(y)$.\footnote{In principle we can also think of non-parametric approaches, but these are beyond the scope of the current paper.} We postpone the characterization of parameter spaces until Section \ref{sec:ambig}.
\end{remark}

\noindent From the last expression in equation (\ref{appeq:mrsint}) we define the integral operator $\mathcal{T}$ as follows:
\begin{equation}
	\label{appeq:mrsintoper}
	\nu = \mathcal{T} \left[ \nu \right] + g.
\end{equation}
\begin{remark}
	Equations (\ref{appeq:mrs}) through (\ref{appeq:mrsintoper}) involve integration of functions $\psi(y,x):$ $\mathbb{R}^{n_y} \times\mathbb{R}^{n_y} $ $\rightarrow$ $\mathbb{R}$,  $f(y|x):$ $\mathbb{R}^{n_y} \times\mathbb{R}^{n_y} $ $\rightarrow$ $\mathbb{R}$, and $\omega(y):$ $\mathbb{R}^{n_y}$ $\rightarrow$ $\mathbb{R}$.  We will refrain here from stating formal assumptions to guarantee the integrals are well defined and the implied integral operator $\mathcal{T}$ is bounded.  We refer to \cite{tauchen1991quadrature} for a formal discussion of regularity conditions.
\end{remark}

\noindent Next, we impose a straightforward (for our particular context) technical assumption regarding $h(y)$, namely since $h_{t+1}$ = $d_{t+1}/d_{t}$ = $h(y_{t+1}),$ we exclude negative dividends with the following assumption:
\begin{guess}
	\label{assum:nonneg}
	The function $h:$ $\mathbb{R}^{n_y}$ $\rightarrow$ $\mathbb{R}$, is non-negative.
\end{guess}
\noindent And because $\psi(y,x)$ = $h(y)$ $\times$ $m(y,x)$ with the SDF being positive by no-arbitrage conditions, we also have by implication of the above assumption that $\psi(y,x)$ $\geq$ 0 $\forall$ $x$ and $y$. 

The integral (operator) equation (\ref{appeq:mrsintoper}) will be approximated by the $N$-point quadrature rule $\nu_N$ = $\mathcal{T}_N\nu_N$ + $b_N$. To that end, let $\bar{y}_k$ and $w_k$ for $k$ = 1, $\ldots$, $N$, be the abscissa and weights for an $N$-point quadrature rule for the density $\omega(y)$. 
A quadrature rule can be viewed as a discrete probability model that approximates the density $\omega$, and in the case of Gauss quadrature rules these approximations are close to minimum norm rules,  yielding for $\psi_{jk}$ = $\psi(\bar{y}_k,\bar{y}_j):$
\begin{eqnarray}
	\label{appeq:linquadeq}
	\bar{\nu}_{Nj} & = & \sum_{k=1}^N \left[1 + \bar{\nu}_{Nk} \right] \psi_{jk} \pi^N_{jk}  \notag \\
	& = & \sum_{k=1}^N  \left[\bar{\nu}_{Nk}  \right] \psi_{jk} \pi^N_{jk} + \sum_{k=1}^N \left[1 \right] \psi_{jk} \pi^N_{jk} \quad j = 1, \ldots, N
\end{eqnarray}
where $\pi^N_{jk}$ = $\pi_k^N(\bar{y}_j)$ for $\pi_k^N(x)$ = $[f(\bar{y}_k|x)/(s(x) \omega(\bar{y}_k))] w_k$ and $s(x)$ = $\sum_{j=1}^{N} [f(\bar{y}_j|x)/\omega(\bar{y}_j)] w_i$. Therefore:
\begin{equation}
	\label{appeq:linquadfinal}
	\underset{\bar{\nu}_{N}}{\underbrace{
			\left[
			\begin{array}{c}
				\bar{\nu}_{N1} \\
				\vdots \\
				\bar{\nu}_{Nj} \\
				\vdots \\
				\bar{\nu}_{NN}
			\end{array}
			\right]}}	 =  
	\underset{\mathcal{T}_N \bar{\nu}_{N}}{\underbrace{\left[
			\begin{array}{cccc}
				\psi_{11} \pi^N_{11} & \ldots & &  \psi_{1N} \pi^N_{1N}	 \\
				& & & \\
				\vdots &  \ddots & & \vdots \\
				& &   & \\
				\psi_{N1} \pi^N_{N1} & \ldots & & \psi_{NN} \pi^N_{NN}
			\end{array}
			\right]
			\left[
			\begin{array}{c}
				\bar{\nu}_{N1} \\
				\vdots \\
				\bar{\nu}_{Nj} \\
				\vdots \\
				\bar{\nu}_{NN}
			\end{array}
			\right]}}
	+ \underset{b_N}{\underbrace{
			\left[
			\begin{array}{c}
				\sum_{k=1}^N \psi_{1k} \pi^N_{1k} \\
				\vdots \\
				\sum_{k=1}^N	 \psi_{jk} \pi^N_{jk} \\
				\vdots \\
				\sum_{k=1}^N	 \psi_{Nk} \pi^N_{Nk}
			\end{array}
			\right]}}.
\end{equation}
\begin{remark}
	Under suitable regularity conditions, see e.g.\ \cite{davis2014methods} and \cite{tauchen1991quadrature}, the approximate solution $\nu_N$ = $\left[ I - \mathcal{T}_N \right]^{-1}b_N$ convergences to the function  $\nu(x)$ $\in$ $\mathbb{R}$ solving the integral equation (\ref{appeq:mrsint}). We refrain here from stating formal assumptions as the main focus of the paper is discrete state approximations. For our analysis we keep $N$ finite for a number of technical reasons. Dealing with infinite dimensional settings is left for future research, as explained in the Conclusion Section \ref{sec:concl}.
\end{remark}

To proceed, recall that  $\psi(y,x)$ = $h(y)$ $\times$ $m(y,x)$, which motivates defining the following matrices $\Psi_N$ := $\left[\psi_{ij}\right]_{i,j =1, \ldots, N}$, $\mathcal{M}_N$ := $\left[m_{ij}\right]_{i,j =1, \ldots, N}$, $\Pi_N$ := $\left[\pi_{ij}\right]_{i,j =1, \ldots, N}$, $\mathcal{H}_N$ = $\left[[h_i]_{i= 1, \ldots, N} \times \mathbf{1}_N^\prime\right]$ where $\mathbf{1}_N$ is a $N$ $\times$ 1 vector of ones, and finally $\Psi_N$ = $\mathcal{H}_N \circ \mathcal{M}_N$. 
\begin{guess}
	\label{appassum:inverse}
	The $N \times N$ matrix $[I_N - \Psi_N \circ \Pi_N]$ is full rank and therefore invertible.
\end{guess}
\noindent Under Assumption \ref{appassum:inverse} the following is an approximate solution to the fundamental pricing equation (\ref{eq:inteq2}):
\begin{equation}
	\label{appeq:discreteinversion1}
	\bar{\nu}_{N} =  [I_N - \Psi_N \circ \Pi_N]^{-1} b_N \quad \text{ or equivalently } \quad  \bar{\nu}_{N} =  [I_N - \mathcal{H}_N \circ \mathcal{M}_N \circ \Pi_N]^{-1} b_N
\end{equation}
which can be viewed as the solution to the asset pricing equations where the law of motion of the state vector is a discrete Markov chain with the $N$ quadrature abscissa $\bar{y}_j$ as states and transition probabilities $\pi^N_{jk}$ = Pr($y_t$ =  $\bar{y}_k$ $\vert$ $y_{t-1}$ = $\bar{y}_j$). Put differently, we can think of the discrete Markov chain as a proxy for  $f(y|x)$ in equation (\ref{appeq:mrsint}) and the Nystrom extension of the solution to the entire domain of $x$ is
\begin{equation}
	\tilde{\nu}_{N}(x)  =  \sum_{k=1}^N \left[1 + \bar{\nu}_{Nk} \right] \psi(\bar{y}_k,x) \pi^N_{k}(x) \qquad x \in \mathbb{R}^M.
\end{equation}
\noindent Inspecting equation (\ref{appeq:linquadfinal}) we can rewrite it as follows:
$\bar{\nu}_{N}$ =  $[\Psi_N \circ \Pi_N] \bar{\nu}_{N}$ + $b_N$, or equivalently $\bar{\nu}_{N}$ =  $[\mathcal{H}_N \circ \mathcal{M}_N \circ \Pi_N] \bar{\nu}_{N}$ + $b_N$. Looking at $\mathcal{H}_N$, $\mathcal{M}_N$ and $\Pi_N$ separately has the advantage that we can distinguish the role of dividend growth from that of the stochastic discount factors and/or the transition probabilities. By the same token we can simply use $\Psi_N$ as the discounted dividend growth process as a joint component.  

\begin{remark}
	In equation (\ref{appeq:mrsint}) we presented a quadrature rule 
	where the weights $w_k$ only depend on an unconditional density $\omega$, as suggested by \cite{tauchen1991quadrature}. 
	\cite{farmer2017discretizing}
	provide a method for accurately discretizing general Markov processes by
	matching low order moments of the conditional distributions using maximum entropy. We use their approach. 
\end{remark}

\noindent The next assumption is needed to rely on the Markov chain convergence theorem, see e.g.\ \cite{meyn1993markov}. 
\begin{guess}
	\label{assum:Harris}
	The discrete state Markov chain characterized by transition density $\Pi_N$  is aperiodic and Harris recurrent.
\end{guess}
\noindent Under Assumption \ref{assum:Harris} we can characterize the ergodic distribution denoted as $\pi^e_N$ as the eigenvector corrresponding to the eigenvalue 1 of the matrix  $\Pi_N$. 

\medskip

\begin{remark}
	So far we focused on dealing with solving equation (\ref{eq:inteq2}) and as we explained in the Introduction, the QC solution methods involve finding eigenvalues and eigenvectors of $ [I_N - \Psi_N \circ \Pi_N]$. It should be noted that much of our analysis also applies to an arguably simpler problem:
	\begin{equation}
		\rho \phi(x_t) = \int m(x_{t+1},x_t) \phi(x_{t+1}) f(x_{t+1}|x_t) dx_{t+1}, \label{appeq:inteq1}
	\end{equation}
	which pertains to  \cite{hansen2009long} who link the permanent-transitory decomposition of stochastic discount factors (SDF) $m(x_{t+1},x_t)$ in Markovian settings, with a $M$-dimensional driving process $x_t$ and the conditional density $f(x_{t+1}|x_t)$, to a Perron-Frobenius eigenfunction problem. The eigenvalue $\rho$ determines the average yield on long-horizon payoffs and the eigenfunction  $\phi(\cdot)$ is positive and characterizes dependence of the price of long-horizon payoffs on the Markov state.
	Using a $N$-point discretization we can use standard Perron-Frobenius theory to solve the eigenvector 
	$\rho_N$ $\phi_N$ = $[\mathcal{M}_N \circ \Pi_N]$ $\phi_N$. More specifically, for a discrete time Markov chain, see Example 6.1 of \cite{hansen2009long}, we can write:
	{\footnotesize
		\begin{equation}
			\label{appeq:LRR1}
			\rho_N 	\left[
			\begin{array}{c}
				\phi_{N1} \\
				\vdots \\
				\phi_{Nj} \\
				\vdots \\
				\phi_{NN}
			\end{array}
			\right] =	\left[
			\begin{array}{cccc}
				m_{11} \pi^N_{11} & \ldots & &  m_{1N} \pi^N_{1N}	 \\
				& & & \\
				\vdots &  \ddots & & \vdots \\
				& &   & \\
				m_{N1} \pi^N_{N1} & \ldots & & m_{NN} \pi^N_{NN}
			\end{array}
			\right]
			\left[
			\begin{array}{c}
				\phi_{N1} \\
				\vdots \\
				\phi_{Nj} \\
				\vdots \\
				\phi_{NN}
			\end{array}
			\right],	 
	\end{equation}}
	where the principal eigenvector characterizes the long-run risk associated with the SDF. No-arbitrage implies $\mathcal{M}_N$ $>$ 0 and for $\Pi_N$ we can make the following observation.
	As noted by \cite{hansen2009long} in their Example 6.1, a principal
	eigenvector is found by finding an eigenvector of  $[\mathcal{M}_N \circ \Pi_N]$ with strictly positive entries.
	Standard Perron-Frobenius theory implies that if the chain is irreducible (cfr.\ Assumption \ref{assum:Harris}),
	since the multiplicative functional is strictly positive, there is such an eigenvector which is unique up to scale.\footnote{The convergence of $\rho_N$ and $\phi_N$ to respectively $\rho$ and $\phi$ in equation  (\ref{appeq:inteq1}) - assuming their existence and uniqueness - is further discussed in \cite{hansen2009long}.}
\end{remark}

\medskip

Finally, we introduce some generic notation for equation (\ref{eq:discreteinversion1}) and a variation of it more suitable for quantum computing. Starting with the former, we will use the following generic notation:
\begin{equation}
	\label{appeq:finalquadeq}
	\mathcal{A}_N	\bar{\nu}_{N} =   b_N \quad \text{ with } \mathcal{A}_N = [I_N - \mathcal{T}_N] = [I_N -  \mathcal{H}_N \circ \mathcal{M}_N \circ \Pi_N]\quad \text{ and therefore } \bar{\nu}_{N} = \mathcal{A}_N^{-1}b_N.
\end{equation}
Quantum mechanics involve unitary operator applications to unit-norm vectors in Hilbert spaces. What that in mind, we will rewrite $b_N$ as:
$b_N$  $\equiv$ diag$(\sqrt{N}\sum_{k=1}^N \psi_{1k} \pi^N_{1k}, \ldots, 
\sqrt{N}\sum_{k=1}^N	 \psi_{Nk} \pi^N_{Nk}) \times \iota_N$ 
$\equiv$  $\mathcal{B}_N \iota_N,$
with $\iota_N$ is a unit-norm vector in a $N$-dimensional space. We can rewrite equation (\ref{appeq:finalquadeq}) as:
\begin{equation}
	\label{appeq:finalquadequnitnorm}
	(\mathcal{B}^{-1}_N\mathcal{A}_N)	\bar{\nu}_{N} =   \iota_N \quad \text{ and therefore } \bar{\nu}_{N} = \mathcal{C}_N^{-1}\iota_N \quad \text{ with } \mathcal{C}_N = (\mathcal{B}^{-1}_N\mathcal{A}_N).
\end{equation}
Note that $\mathcal{C}_N$ = $\mathcal{B}^{-1}_N\mathcal{A}_N$ encodes all the information about the model and we can therefore focus on $\mathcal{C}_N$ when we study multiple models.

\subsection{Dividend Growth Model}

We start with equation (2) of \cite{hansen2008consumption} applied to log dividend growth instead of consumption. We download data from 1964Q1 to 2020Q4 from FRED (Personal income receipts on assets: Personal dividend income (series: B703RC1Q027SBEA), which is seasonally adjusted) and then deflate using CPI growth. Then log real dividend growth is modeled as follows:
\begin{eqnarray}
	\label{appeq:logdivgrowth}
	d_{t+1} - d_t &=& a+x_{t+1}+be_{t+1}\\
	x_{t+1} &=& \rho_1 x_{t}+c \epsilon_{t+1},  \notag
\end{eqnarray}
where the shocks $e_{t+1} \sim N(0,1)$ and $\epsilon_{t+1} \sim N(0,1)$.

\begin{figure}[h]
	\caption{Observations and predictions with 95\% confidence intervals}
	\label{graph:Div}
	\includegraphics[scale=0.6]{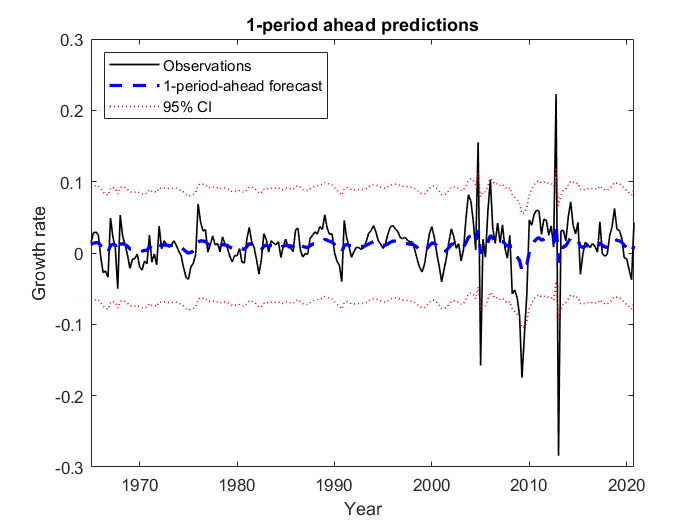}
	\centering
\end{figure}	
We collect the parameters of the above model into the vector $\theta_L$ = $(a, b, c, \rho_1)^\prime$.
We can estimate the parameters of the model (\ref{appeq:logdivgrowth}) via Maximum Likelihood. To estimate the model we use data covering the sample 1964Q1 to 2020Q4 from FRED (Personal income receipts on assets: Personal dividend income (series: B703RC1Q027SBEA), which is seasonally adjusted) and then deflate using CPI growth. 
Table \ref{tab:Solutions_div} reports the parameter estimates and their standard errors. Note that all the parameters are statistically significant. The one-period ahead mean predictions appear in Figure \ref{graph:Div}. During the sample we observe some extreme observations, mostly related to the financial crisis, which are not well captured by the model.
\begin{table}[h]
	\centering
	\caption{Parameter estimates for log real dividend growth}
	\label{tab:Solutions_div}
	\begin{tabular}{c c c}
		Parameter&Coefficient&Standard Error \\  
		\hline 
		$\rho_1$ &0.64079&0.25901
		\\  
		$c$ &0.01520&0.00901
		\\  
		$b$ &0.03630&0.00272
		\\  
		$a$ &0.01037&0.00416
		\\  
	\end{tabular}
\end{table}

\begin{figure}[h]
	\caption{Distribution of Kullback-Leibler divergence from asymptotic distribution random draws}
	\label{graph:KL_dist}
	\includegraphics[scale=0.5]{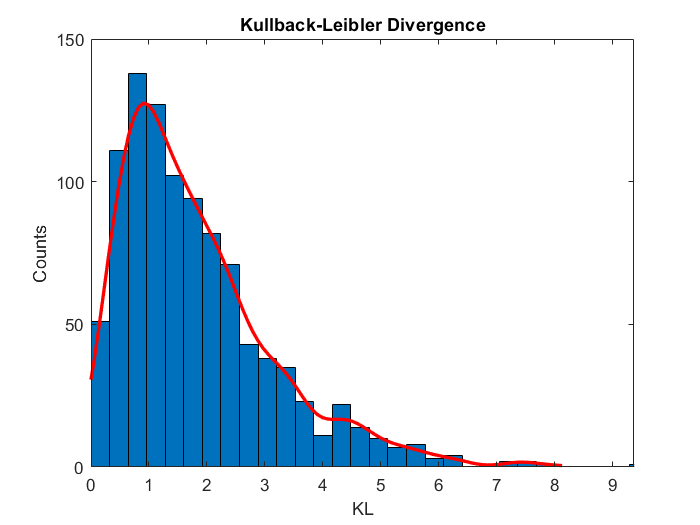}
	\centering
\end{figure}

We rely on the asymptotic distribution theory for MLE to characterize model uncertainty about equation (\ref{appeq:logdivgrowth}). Namely, we assume that the model is correctly specified, but we are uncertain about its parameter values. We create one thousand draws of parameter models $\{\theta_L^i, i = 1, \ldots 1000 \}$ and construct a 1000 models for log dividend growth. More specifically, we draw one thousand $\theta_L^i \sim N(\hat{\theta}_L, \hat{\Sigma})$,  excluding draws with $\rho_1 < 0$ and $\rho_1 > 1$. The collection of models therefore represents statistical uncertainty, since we assume the model specification is correct and asymptotic distribution theory provides us guidance about parameter uncertainty.
We can characterize and visualize the uncertainty we calculate the
Kullback-Leibler divergence $KL_i=\frac{1}{2}[(\theta_L^i - \hat{\theta}_L)' \hat{\Sigma}^{-1}(\theta_L^i - \hat{\theta}_L)]$. The distribution appears in Figure \ref{graph:KL_dist}.

\medskip

\subsection{Discretizations for CRRA and Recursive Utility}

Next, we specify the log of the one period stochastic discount factor $s_{t+1,t}:$
\begin{equation}
	\label{appeq:logSDF}
	s_{t+1,t} = \alpha_0 + \alpha_1 x_t + \xi w_{t+1}.
\end{equation}
We have for the risk-free rate $r_t^f:$
$1/r_t^f$ = $\mathbb{E}_t \exp{[\alpha_0 + \alpha_1 x_t + \xi w_{t+1}]}$ = $\exp{[\alpha_0 + \alpha_1 x_t]} \mathbb{E}_t \exp{[\xi w_{t+1}]}$ = $\exp{[\alpha_0 + \alpha_1 x_t]}$
since $w_{t+1}$ is i.i.d.\ with mean zero. Assuming $r^f_t$ $>$ 0 $\forall$ $t$ (implied by no arbitrage) and taking a sample average, we get:
\[
T^{-1}\sum_{t=1}^T -\log r_t^f = T^{-1}\sum_{t=1}^T  [\alpha_0 + \alpha_1 x_t] = \hat \alpha_0,
\]
since $x_t$ is a mean zero process. This yields an estimate for $\alpha_0$. We use sample average of the risk-free rate - 3-month T-Bill from same sample period of 1964Q1 to 2020Q4.\footnote{Data is downloaded from FRED (\url{https://fred.stlouisfed.org/series/DTB3}) in percentage terms, and we take quarterly average of all daily yield.}
To obtain an estimate for $\alpha_1$ we note that:
\[
-\log r_t^f - \hat \alpha_0 = \alpha_1 x_t = \alpha_1 [\hat x_t - \epsilon_t]
\]
where $\hat x_t$ are the Kalman filtered state estimates with $\hat x_t$ $\perp$ $\epsilon_t$, where $\epsilon_t$ is the filtering error.\footnote{The Kalman filter is implemented in MATLAB. The initial state vector and covariance matrix are both $0$.} This means that we can recover an estimate $\hat \alpha_1$ from running a regression of demeaned minus log risk-free rates onto filtered $\hat x_t$. 
Given our sample, the numerical values are as follows: $\hat \alpha_0$ = -0.8974, $\hat \alpha_1$  = 1.2038.
Next, we compute with the implied value with $\beta$ = 0.99 the following (using the formula middle of page 264 of \cite{hansen2008consumption}):
{\footnotesize
	\begin{eqnarray*}
		\xi & = & -\gamma b + (1 - \gamma) \beta [1 \, \, 0 \, \, 0 \, \, 0] 
		\left[ I - \beta \times \left[
		\begin{array}{cccc}
			\rho_1 & \rho_2 &  \rho_3  &  \rho_4 \\
			1 & 0 & 0 & 0 \\
			0 & 1 & 0 & 0 \\
			0 & 0 &  1 & 0
		\end{array}
		\right]
		\right]^{-1}  \left[
		\begin{array}{c}
			c \\ 0 \\ 0 \\ 0
		\end{array}
		\right] \\
		& = & -\gamma b + (1 - \gamma) \beta [1 \, \, 0 \, \, 0 \, \, 0] 
		\left[
		\begin{array}{cccc}
			1 - \beta \rho_1 & - \beta \rho_2 & - \beta \rho_3  & - \beta \rho_4 \\
			- \beta & 1 & 0 & 0 \\
			0 & - \beta & 1 & 0 \\
			0 & 0 &  - \beta & 1
		\end{array}
		\right]^{-1}  \left[
		\begin{array}{c}
			c \\ 0 \\ 0 \\ 0
		\end{array}
		\right]=  -\gamma b + \frac{(1-\gamma)c\beta }{ 1 - \sum_{i=1}^{4} \beta^i \rho_i } 
	\end{eqnarray*}
}
We have estimates for $b$, $c$, $\beta$, and $\rho_i$, so that we can compute $\xi$ as a function of $\gamma:$
\begin{eqnarray*}
	\xi &=&  \frac{c\beta}{ 1 - \beta \rho_1 } -  \left( \frac{c\beta}{ 1 - \beta \rho_1 }+b\right)\gamma \\
	&=& 0.0412 - 0.0775\gamma \qquad \text{Using estimated parameter values.}
\end{eqnarray*}


Next we provide the details of the practical implementation for a single model, which means we select a given $\theta$ = ($\theta_L$, $\gamma$). We set $N$ = 4, or more precisely we compute the quadrature approximation to the AR(1) model for the log dividend growth appearing in equation (\ref{appeq:logdivgrowth}) with $4$ abscissa using the code of \cite{farmer2017discretizing} matching the first two sample moments of the data.\footnote{The code is available at \url{https://github.com/alexisakira/discretization}\, .} This yields the transition matrix  $\Pi_{4}^x(\theta_L)$, using the notation of equation (\ref{eq:discreteinversion1}). The log of the SDF in equation (\ref{appeq:logSDF}) involves a standard Gaussian shock $w_{t+1}$. We assume a one-standard deviation positive/negative shock, i.e.\ $w_{t+1}$ = $\pm$ 1, reflecting a ``good'' versus ``bad'' economic environment. Since the shock is assumed to be independent of $x_t$, we have the following transition density: $\Pi_{8}(\theta)$ = $\Pi_{4}^x(\theta_L) \otimes_K S_2$, where $\otimes_K$ is a Kronecker product and $S_2$ is the two-dimensional matrix where all entries are equal to 1/2.
Recall that we need to invert the matrix $\mathcal{C}_{8}(\theta)$ = $\mathcal{B}^{-1}_{8}(\theta)\mathcal{A}_{8}(\theta)$ where:
\begin{itemize}
	\item $\mathcal{A}_{8}(\theta)$ =  $[I_{8} -  \mathcal{H}_{8}(\theta_L) \circ \mathcal{M}_{8}(\theta) \circ \Pi_{8}(\theta_L)]$
	\item $\mathcal{B}_{8}(\theta)$ = diag($\sqrt{16}\sum_{k=1}^{16} \psi_{1k} \pi_{1k}, \ldots, 
	\sqrt{{16}}\sum_{k=1}^{16}	 \psi_{{16}k} \pi_{{16}k}$).
\end{itemize}
To that end, we first define the following objects:
\begin{itemize}
	\item $x_i(\theta_L)$ for $i$ = 1, $\ldots$, 8 are the abscissa for the quadrature discretization of the AR(1) model
	\item As already noted,  $\Pi_{8}(\theta_L)$ = $\left[\pi_{ij}\right]_{i,j =1, \ldots, {16}}$ = $\Pi_{4}(\theta_L) \otimes_K S_2$.
	\item $\mathbb{S}_{4}(\theta_L)$ = $\left[\exp\left(-0.8970 + 1.2038 x_i(\theta_L)\right)\right]_{i =1, \ldots, 8}$, a $8 \times 1$ vector, using equation (\ref{appeq:logSDF})
	\item  $\mathcal{M}_{8}(\theta)$ = $\left[m_{ij}\right]_{i,j =1, \ldots, 16}$ = $\left[\mathbb{S}_{4}(\theta_L) \mathbf{1}_{4}^\prime\right] \otimes_K \left(\begin{array}{cc}
		\exp(\xi) & \exp(-\xi) \\
		\exp(\xi) & \exp(-\xi) 
	\end{array}\right)$ where $\mathbf{1}_{4}$ is a $8$ $\times$ 1 vector of ones. Note that for the CRRA utility function $\xi$ = $-b\gamma$ = $-0.03630\gamma$, using the parameter estimates reported in Table \ref{tab:Solutions_div}, and $\xi$ =  $0.0412$ - $0.0775\gamma$ for recursive utility with the intertemporal elasticity of substitution  equal to one.
	\item We set $\gamma$ = 2, 5 and 10, which yields a total of six SDF specifications.
	\item Using equation (\ref{appeq:logdivgrowth}) we construct dividend growth:
	$\mathbb{D}_{4}(\theta_L)$ = $\left[\exp\left(0 .01037 +  x_i(\theta_L)\right)\right]_{i =1, \ldots, 8}$, and compute  $\mathcal{H}_{8}(\theta_L)$ = $\left[\mathbb{D}_{4}(\theta_L) \mathbf{1}_{4}^\prime\right] \otimes_K \left(\begin{array}{cc}
		\exp(0.03630) & \exp(-0.03630) \\
		\exp(0.03630) & \exp(-0.03630) 
	\end{array}\right)$ 
	\item  $\left[\psi_{ij}\right]_{i,j =1, \ldots, {16}}$ = $\mathcal{H}_{8}(\theta_L) \circ \mathcal{M}_{8}(\theta)$
\end{itemize} 
With the above, we have all the elements to compute $\mathcal{B}^{-1}_{8}(\theta)$ and $\mathcal{A}_{8}(\theta)$.

\subsection{Models with SV \label{appsubsec:SV}}

We provide further details of the models with SV discussed in Section \ref{subsec:SV}. The SV setup involves a two state model, for convenience called the good economy $G$ and the other the bad economy $B$ state with transition matrix appearing in equation (\ref{eq:transition}) and implied ergodic distribution: $[\pi_G, 1 - \pi_G]$ where $\pi_G$ = $p_{BG}/(p_{BG} + p_{GB}).$  The QMLE interpretation in Section \ref{subsec:SV} allows us to anchor one of the two regime volatilities since: $b$ = $\pi_G \times b_G$ + $(1 - \pi_G)b_B,$ with $b_B$ $>$ $b$ $>$ $b_G,$. We define the good regime volatility with the parameter $\gamma_G$ = $b_G / b$.  Given $\pi_G$ and $\gamma_G$, we can show that $b_B$ = $b(1 - \pi_G \gamma_G)/(1 - \pi_G).$ As a result, we only need to fix $\pi_G$ and $\gamma_G$ to study models with SV. 

\medskip

Since the shock is assumed to be independent of $x_t$, we have the following transition density: $\Pi_{8}(\theta)$ = $\Pi_{4}^x(\theta_L) \otimes_K S_2(\pi_G)$, where $S_2(\pi_G)$ is the two-dimensional matrix $[\pi_G, 1 - \pi_G] \otimes_K \mathbf{1}_{2}.$ Having specified the transition matrix, now dependent not only on $\theta_L,$ but also $\pi_G,$ we proceed to modifying the dividend growth with SV as follows:
$\mathbb{D}_{4}(\theta_L)$ = $\left[\exp\left(0 .01037 +  x_i(\theta_L)\right)\right]_{i =1, \ldots, 8}$, and:  
\begin{equation}
	\label{appeq:divgrowthSV}
	\mathcal{H}_{8}(\theta_L) = \left[\mathbb{D}_{4}(\theta_L) \mathbf{1}_{4}^\prime\right] \otimes_K \left(\begin{array}{cc}
		\exp(b_G) & \exp(-b_G) \\
		\exp(b_B) & \exp(-b_B) 
	\end{array}\right).
\end{equation}
Regarding the SDF, we modify the model as follows:
\[
\mathcal{M}_{8}(\theta) = \left[m_{ij}\right]_{i,j =1, \ldots, 16} = \left[\mathbb{S}_{4}(\theta_L) \mathbf{1}_{4}^\prime\right] \otimes_K \left(\begin{array}{cc}
	\exp(\xi_{G}) & \exp(-\xi_{G}) \\
	\exp(\xi_{B}) & \exp(-\xi_{B}) 
\end{array}\right)
\]
with: $\xi_{G}$ = $-b_G\gamma$ and $\xi_{B}$ = $-b_B\gamma$ for CRRA. For the IES model we have and $\xi_{G}$ =  $c\beta/(1 - \beta \rho_1) -  \left( c\beta/( 1 - \beta \rho_1)+b_G\right)\gamma$ and similarly for $\xi_B.$ The remaining of the model computations are the same as in the previous subsection.

\subsection{Discretization of Rare Disaster Model \label{appsec:raredis}}

We consider the rare disaster model appearing in equations (\ref{eq:aprare1})-(\ref{eq:apprare2}). 
We follow \cite{farmer2017discretizing} to solve for the equilibrium price–dividend ratio. Suppose the state space of resilience $H_t$ is discretized, and let $k$ = 1, $\ldots,$ $N$ be the states. Since the disaster probability
is constant, it follows that (using the notation of equation (\ref{eq:linquadfinal})):
\[
\bar{\nu}_{Nj} = 1 + \exp{(-\delta + g_D)} (1 + h_j) \sum_{k=1}^N \pi_{jk}^N	\bar{\nu}_{Nk}
\]
where $\bar{\nu}_{Nj}$ is the price–dividend ratio in state $j,$ $h_j$ is the resilience in state $j,$ and $\pi_{jk}^N$ is the transition probability from state $j$ to $k.$ Then for $\bar{\nu}_{N}$ the discretized price-dividend ratio and $h_N$ = ($h_{N1}$ $\ldots,$ $h_{NN}$) be resilience in each state, we have
\[
\bar{\nu}_{N} =  1 + \exp{(-\delta + g_D)} \text{diag}(1+h)\Pi_N\bar{\nu}_{N} 
\]
with $\Pi_N$ the transition matrix, yielding a discrete approximation to the model closed-form solution. For the parameter values, following \cite{gabaix2012variable} we set the discount rate $\delta$ =
0.0657, relative risk aversion $\gamma$ = 4, the dividend growth rate $g_D$ =
0.025, disaster probability $p$ = 0.0363, the consumption recovery rate $B$ = 0.66, and the speed of mean reversion $\phi_H$ = 0.13. The implied value for the constant $H_*$ equals 0.09.
Because the resilience process is highly nonlinear, we need many grid points to obtain an accurate solution. Calculations reported by \cite{farmer2017discretizing} suggest for practical purposes $N$ = 11 is adequate.

\subsection{HHL Feasibility
	\label{appsec:hhl-feasibility}}

The exponential speedup HHL offers with respect to the system size is offset by the runtime dependence on sparsity and conditioning. Therefore, HHL is a promising approach only for applications when $A$ is $s$ sparse and well conditioned. For models that do not fit this criteria, quantum alternatives to HHL with better dependence on $s$ and $\kappa$ are available (see \cite{wossnig2018quantum}, \cite{ambainis2010variable}). In this context, the sparsity is measured by the largest number of entries in a single row that have an absolute value greater that $1\mathrm{e}{-5}$. In Table \ref{tab:sparsecondition} we show the values of $s$ and $\kappa$ for the ten model specifications we study with $N$ = 64 abscissa. Recall that a model with $N$ = 64 results in an 256x256 $A$ matrix to be inverted. We find that the sparsity of the matrix used to solve the models with and without stochastic volatility does not grow linearly with $N$. The rare disasters model requires inverting a matrix that is poorly conditioned, which makes the quantum approach less appealing for this class of models. While there is no exact cutoff for the condition and sparsity that determines when HHL is practical for a given problem, these results lead us to believe that the proposed method for the models with and without stochastic volatility are suitable for a quantum linear algebra algorithm on a fault tolerant quantum processor.

\begin{table}[h]
	\centering
	\caption{Sparsity (a) and condition number (b) of the $A$ matrix produced by different model specifications with 64 abscissa.}
	\label{tab:sparsecondition}
	\medspace
	\begin{subtable}{0.45\textwidth}
		\centering
		\medspace
		\caption{The sparsity $s$ of the $A$ matrix inverted in the asset pricing model with the listed utility functions and $\gamma$ specifications.}
		\begin{tabular}{l c c c}
			Models & Constant & $\pi_G$=0.8 & $\pi_G$ = 0.95 \\  
			& Volatility & $\gamma_G$ = 0.3 & $\gamma_G$ = 0.01 \\  
			\hline 
			CRRA & 82 & 82 & 82 \\  
			IES = 1, & 80 & 82 & 82 \\
			$\gamma$ = 2 & & & \\  
			IES = 1, & 82 & 82 & 82 \\  
			$\gamma$ = 10 & & & \\  
			RD & 78 & - & - \\
		\end{tabular}	
	\end{subtable}
	\hfill
	\begin{subtable}[t]{0.45\textwidth}
		\centering
		\caption{The condition number $\kappa$ of the $A$ matrix inverted in the asset pricing model with the listed utility functions and $\gamma$ specifications.}
		\begin{tabular}{l c c c}
			Models & Constant & $\pi_G$=0.8 & $\pi_G$ = 0.95 \\  
			& Volatility & $\gamma_G$ = 0.3 & $\gamma_G$ = 0.01 \\  
			\hline 
			CRRA & 4.39 & 4.10 & 5.88 \\  
			IES = 1, & 3.89 & 3.99 & 2.95 \\ 
			$\gamma$ = 2 & & & \\   
			IES = 1, & 6.91 & 6.23 & 5.59 \\
			$\gamma$ = 10 & & & \\    
			RD & 82.9 & - & - \\
			
		\end{tabular}
	\end{subtable}
\end{table}

Table \ref{tab:depthanderror} features the circuit depth and ideal fidelity of a simulated 16 $\times$ 16 HHL implementation using the listed model parameters with 4 ancillary qubits. The circuit depth is a useful metric to estimate the hardware noise we can expect when running a circuit on a quantum processor. We calculate the depth as defined by \textit{Qiskit} on the IBM Torino Heron r1 processor. Circuits of this depth cannot be accurately executed on current processors, however hardware and error correction advances could make them realistic in the near future. On a fault tolerant quantum processor, the solution fidelity will approach the noiseless estimates in panel (a). We could increase the ideal fidelity by using more ancillary bits for Quantum Phase Estimation within the HHL circuit at the cost of a greater circuit depth. 

\begin{table}[h]
	\centering
	\caption{The fidelity (a) and circuit depth (b) of a naive HHL implementation on a noiseless simulator with all-to-all connectivity. All sum produced as a result of the asset pricing model with the listed utility function and $\gamma$ specifications.}
	\label{tab:depthanderror}
	\medskip
	\begin{subtable}{0.45\textwidth}
		\centering
		\caption{The solution fidelity of an HHL result on an all-to-all connectivity processor that could be used to solve a model with the listed utility function and $\gamma$ specifications.}
		\begin{tabular}{l c c c}
			Models & Constant & $\pi_G$=0.8 & $\pi_G$ = 0.95 \\  
			& Volatility & $\gamma_G$ = 0.3 & $\gamma_G$ = 0.01 \\  
			\hline 
			CRRA & 0.90 & 0.92 & 0.94 \\  
			IES = 1, & 0.91 & 0.91 & 0.93 \\ 
			$\gamma$ = 2 & & & \\   
			IES = 1, & 0.97 & 0.88 & 0.91 \\
			$\gamma$ = 10 & & & \\    
			RD & 0.32 & - & - \\
			
		\end{tabular}
	\end{subtable}
	\hfill
	\begin{subtable}{0.45\textwidth}
		\centering
		\caption{The depth of an HHL circuit on an all-to-all connectivity processor that could be used to solve a model with the listed utility function and $\gamma$ specifications.}
		\begin{tabular}{l c c c}
			Models & Constant & $\pi_G$=0.8 & $\pi_G$ = 0.95 \\  
			& Volatility & $\gamma_G$ = 0.3 & $\gamma_G$ = 0.01 \\  
			\hline 
			CRRA & 40268 & 39512 & 39915 \\  
			IES = 1, & 39758 & 39774 & 39596 \\ 
			$\gamma$ = 2 & & & \\   
			IES = 1, & 40023 & 39892 & 39924 \\
			$\gamma$ = 10 & & & \\    
			RD & 39321 & - & - \\
			
		\end{tabular}
	\end{subtable}
\end{table}

\end{document}